\documentclass[twocolumn]{aastex63}

\usepackage{graphicx}	
\usepackage{amsmath}	
\usepackage{amssymb}	
\usepackage{xcolor}
\usepackage{booktabs}   

\usepackage{graphicx}
\usepackage{amssymb}
\usepackage{amsmath}
\usepackage{natbib}
\usepackage{wrapfig}
\usepackage[utf8]{inputenc}
\usepackage[T1]{fontenc}
\usepackage{url}
\PassOptionsToPackage{hyphens}{url}\usepackage{hyperref}
\hypersetup{breaklinks=true}
\usepackage{hyperref}
\usepackage[letterspace=0]{microtype}

\tabletypesize{\footnotesize}

\newcommand{\G}{G_{mag}}

\newcommand{\masyr}{ \ {\rm{mas \ yr^{-1}}}\>}
\newcommand{\uasyr}{\ \mu{\rm as}\,{\rm yr}^{-1}}
\newcommand{\mas}{ \ {\rm{mas}}\>}

\newcommand{\kms}{ \ {\rm{km \ s^{-1}}}\>}

\newcommand{\foo}[1]{}

\newcommand{\kpc}{\>{\rm kpc}}

\newcommand{\degree}{\degr}

\newcommand{\pmra}{\mu_{\alpha\star}}
\newcommand{\pmdec}{\mu_{\delta}}

\newcommand{\vlos}{v_{\mathrm{LOS}}}

\newcommand{\hyperfootnote}[1][]{\def\ArgI\hyperfootnoteRelay}
\newcommand\hyperfootnoteRelay[2][]{\href{#1#2}{\ArgI}\footnote{\href{#1#2}{#2}}}

\newcommand{\svlos}{\sigma(v_{\mathrm{LOS}})}
\newcommand{\svt}{\sigma(v_{\mathrm{T}})}
\newcommand{\svr}{\sigma(v_{\mathrm{R}})}
\newcommand{\dsun}{d_{\odot}}
\newcommand{\spmt}{\sigma(\mu_{\mathrm{T}})}
\newcommand{\spmr}{\sigma(\mu_{\mathrm{R}})}
\newcommand{\gaiahub}{\textsc{GaiaHub}}
\newcommand{\code}[1]{\texttt{#1}}

\newcommand{\Gaia} {\textit{Gaia}}
\newcommand{\HST} {\textit{HST}}
\newcommand{\HSTONEPASS} {\textit{HST1PASS}}

\submitjournal{ApJ}

\shorttitle{\gaiahub: proper motions combining HST and Gaia}
\shortauthors{del Pino et al.}
\graphicspath{{./}{figures/}}

\begin{document}

\title{\lsstyle
\Large{\gaiahub: A method for combining data from the Gaia and Hubble space telescopes to derive improved proper motions for faint stars}}

\correspondingauthor{Andres del Pino}
\email{adelpino@cefca.es}

\author[0000-0003-4922-5131]{Andr\'es del Pino}
\affiliation{Centro de Estudios de F\'isica del Cosmos de Arag\'on (CEFCA), Unidad Asociada al CSIC, Plaza San Juan 1, 44001, Teruel, Spain}
\affiliation{Space Telescope Science Institute, 3700 San Martin Drive, Baltimore, MD 21218, USA}

\author[0000-0001-9673-7397]{Mattia Libralato}
\affiliation{AURA for the European Space Agency (ESA), ESA Office, Space Telescope Science Institute, 3700 San Martin Drive, Baltimore, MD 21218, USA}

\author[0000-0001-7827-7825]{Roeland P. van der Marel}
\affiliation{Space Telescope Science Institute, 3700 San Martin Drive, Baltimore, MD 21218, USA}
\affiliation{Center for Astrophysical Sciences, Department of Physics \& Astronomy, Johns Hopkins University, Baltimore, MD 21218, USA}

\author[0000-0001-8354-7279]{Paul Bennet}
\affiliation{Space Telescope Science Institute, 3700 San Martin Drive, Baltimore, MD 21218, USA}

\author[0000-0003-4207-3788]{Mark A. Fardal}
\affiliation{Space Telescope Science Institute, 3700 San Martin Drive, Baltimore, MD 21218, USA}

\author{Jay Anderson}
\affiliation{Space Telescope Science Institute, 3700 San Martin Drive, Baltimore, MD 21218, USA}

\author[0000-0003-3858-637X]{Andrea Bellini}
\affiliation{Space Telescope Science Institute, 3700 San Martin Drive, Baltimore, MD 21218, USA}

\author[0000-0001-8368-0221]{Sangmo Tony Sohn}
\affiliation{Space Telescope Science Institute, 3700 San Martin Drive, Baltimore, MD 21218, USA}

\author[0000-0002-1343-134X]{Laura L. Watkins}
\affiliation{AURA for the European Space Agency (ESA), ESA Office, Space Telescope Science Institute, 3700 San Martin Drive, Baltimore, MD 21218, USA}

\begin{abstract}
We present \gaiahub, a publicly available tool that combines \Gaia{} measurements with \textit{Hubble Space Telescope} (\textit{HST}) archival images to derive proper motions (PMs). It increases the scientific impact of both observatories beyond their individual capabilities. \Gaia{}  provides PMs across the whole sky, but the limited mirror size and time baseline restrict the best PM performance to relatively bright stars. \textit{HST} can measure accurate PMs for much fainter stars over a small field, but this requires two epochs of observation which are not always available.\gaiahub{} yields considerably improved PM accuracy compared to \textit{Gaia}-only measurements, especially for faint sources $(G \gtrsim 18)$, requiring only a single epoch of \textit{HST} data observed more than $\sim 7$ years ago (before 2012). This provides considerable scientific value especially for dynamical studies of stellar systems or structures in and beyond the Milky Way (MW) halo, for which the member stars are generally faint. To illustrate the capabilities and demonstrate the accuracy of \gaiahub, we apply it to samples of MW globular clusters (GCs) and classical dwarf spheroidal (dSph) satellite galaxies. This allows us, e.g., to measure the velocity dispersions in the plane of the sky for objects out to and beyond $\sim 100\kpc$. We find, on average, mild radial velocity anisotropy in GCs, consistent with existing results for more nearby samples. We observe a correlation between the internal kinematics of the clusters and their ellipticity, with more isotropic clusters being, on average, more round. Our results also support previous findings that Draco and Sculptor dSph galaxies appear to be radially anisotropic systems. 
\end{abstract}




\section{Introduction}\label{sec:intro}

Proper motions from space telescopes have had a significant impact on our understanding of the Milky Way (MW) system and its constituent parts. Historically, PMs have been much more difficult to acquire than radial velocities. This has blurred different MW structural components together, and has encouraged the adoption of plausible assumptions (such as equilibrium, isotropy, or orbit circularity) that have, in some cases, turned out to be incorrect. Acquiring data for the missing two dimensions of velocity has led to a more highly structured, dynamic, and interesting picture of the MW and its satellite system.

The era of space astrometry began in earnest with the \textit{Hipparcos} telescope, but this dedicated astrometric mission was limited to bright nearby stars. Thus for many years the main tool for precise astrometry has been the \textit{Hubble Space Telescope} (\textit{HST}). 
Astrometric results from \textit{HST} include precise motions of the Magellanic clouds
(\citealp{Kallivayalil2006}; \citealp{Piatek2008}; \citealp{Marel2014}), 
suggesting a first infall \citep{Besla2007}; limits on the black hole content of globular clusters \citep{Anderson2010, Haeberle2021}; estimation of globular cluster and dwarf galaxy orbits, with consequent insight into their associations and the MW halo mass 
(\citealp{Milone2006}; \citealp{Piatek2007}; \citealp{Sohn2018});
and measurement of a nearly head-on approach of M31 toward the MW \citep{Sohn2012, Marel2012}.
These PM results required two or more epochs of well-spaced observations, which are not always available. 

The \Gaia{} mission, even with its interim data releases, has now induced an avalanche of scientific results based on PMs of MW stars. This mission reaches 7 magnitudes fainter than its predecessor \textit{Hipparcos}, and thus can probe stars to the outer reaches of the MW halo and beyond. 
By this point \Gaia{} has been used to measure the systemic motions of almost all Milky Way globular clusters \citep[e.g.][]{Helmi2018, Baumgardt2019, Vasiliev2021}
and dwarf galaxies \citep[e.g.][]{Fritz2018, McConnachie2020, Battaglia2022}.
Other notable results include the discovery of the Gaia-Enceladus merger remnant \citep{Belokurov2018, Helmi2018b}, the association of numerous dwarf galaxies with the Magellanic group \citep{Kallivayalil2018} coherent motions in the MW disk \citep{Antoja2018}, and the repeated impact of the Sagittarius dwarf galaxy in the star formation of the MW \citep{Ruiz-Lara2020}.

In the most recent \Gaia{} data release, Early Data Release 3 (EDR3), the observation interval is only 34 months. This limits the PM precision that can be obtained using \Gaia{} alone. The time baseline will increase in later data releases, of course, with a maximum possible mission length of around 10 years. However, future releases of PMs are expected to be infrequent and their arrival several years away, in part due to the huge processing task involved in creating each astrometric data release. \Gaia{} also does not image the sky nearly as deeply as is typical with \textit{HST}, due both to the difference in mirror sizes and the contrast between the sky-scanning and static-pointing observing methods of the two telescopes. Hence \textit{Gaia}'s astrometric errors rise rapidly from $G \sim 12$ down to its catalog limit of $ G \sim 21$. A star at this limiting magnitude has a PM uncertainty of 1.4~mas yr$^{-1}$, which for a star at a distance $\dsun =50 \kpc$ translates to a tangential velocity uncertainty of 330~km~s$^{-1}$, too large to be of much use. Even the brightest stars in the Sculptor galaxy ($\dsun = 86 \kpc$), for example, have a tangential velocity error of $\sim 25 \kms $, which is much larger than the true internal dispersion of that galaxy \citep{Martinez-Garcia2021}.

In principle, combining \Gaia{} with older \textit{HST} observations can yield more precise PMs by providing a longer time baseline. \textit{HST} has only surveyed a small fraction of the sky, but even so it has targeted many objects of interest. Many of these \textit{HST} observations date to 10--15 years before the launch of \textit{Gaia}, extending the current time baseline of \Gaia{} by a factor of 4--6. This general idea was used by \citet{Massari2017, Massari2018, Massari2020} to measure PMs of stars in the Draco and Sculptor dSph galaxies, as well as the globular cluster NGC\ 2419. There are other scientific targets that could benefit from a similar approach. However, learning about the data characteristics and archival formats of two different space missions can be a daunting task, which may deter some potential users from attempting to combine \Gaia{} and \textit{HST} data.

In this paper we present \gaiahub, a new code package which measures PMs from the combination of \Gaia{} and \textit{HST} data. The code is designed to be run at a wide variety of user input levels. It is capable of running nearly automatically once given a target or allows extensive customization through optional keywords for greater control over the process by the user. The code will automatically discover, download, and analyze \textit{HST} images in a sky region of interest, and combines the results with the \Gaia{} catalog to produce accurate PMs. The code is published on Zenodo\footnote{\url{https://doi.org/10.5281/zenodo.6467326}} \citep{GaiaHub_zenodo} and Github\footnote{\url{https://github.com/AndresdPM/GaiaHub}} and is free to use and modify.

In this paper we first discuss the general method implemented in \gaiahub{} (\S~\ref{sec:How_it_works}). We then present some science demonstration cases (\S~\ref{sec:Results}), including the internal kinematics of both GCs and dSph galaxies. While statistical errors are estimated automatically by \gaiahub, we mention some systematic errors originating from both \textit{HST} and Gaia that are not included in these estimates (\S~\ref{sec:Systematics}). We then discuss which types of problems are likely to benefit from the use of combined Gaia-\textit{HST} data (\S~\ref{sec:Usability}). The last section summarizes our work (\S~\ref{sec:Conclusions}). An overview of how to use the code, with specific calling sequences and discussion of some of the main options can be found in the Appendix~\ref{Apx:GaiaHub_Execution}.

\section{\gaiahub{} Under the Hood: Combining HST and Gaia}\label{sec:How_it_works}

\gaiahub{} is a software tool that derives PMs by comparing the position of the stars measured with Gaia, $r_{\rm Gaia} = (\alpha, \delta)_{\rm Gaia}$, with those measured by \textit{HST}, $r_{\rm HST} = (\alpha, \delta)_{\rm HST}$. For faint stars ($G \gtrsim 17$), the longer time baseline afforded by combining \textit{HST} and \Gaia{} provides higher precision than is possible using \Gaia{} alone. In this section we describe the technique in detail. In short, we establish a common absolute reference frame between \textit{HST} observations using \Gaia{} stars. This allows PM measurements in any \textit{HST} field without the need to use background galaxies as a reference system.

\subsection{Precise Astrometry with {HST}}\label{sec:HST_astrometry}

High-precision astrometry with \textit{HST} data is instrumental to obtain the necessary PM precision. \gaiahub{} performs the \textit{HST} data reduction in a similar way to that described in other PM analyses based on \textit{HST} data \citep[e.g.][]{Bellini2017, Bellini2018, Libralato2018, Libralato2019}.

\gaiahub{} can work with ACS/WFC and WFC3/UVIS \texttt{\_flc} images, as well as ACS/HRC \texttt{\_flt} exposures. Both \texttt{\_flt} and \texttt{\_flc} images preserve the unresampled pixel data for optimal PSF fitting. WFPC2 and WFC3/IR data are not supported, because the astrometric precision reachable with these data is generally too low for \gaiahub's expected use cases.

The position and flux of all the detectable sources in the field are extracted via empirical point-spread function (PSF) fitting using \HSTONEPASS\footnote{\url{https://www.stsci.edu/~jayander/HST1PASS/}}. For each image, the code fine-tunes the library PSFs taking into account the spatial position on the \textit{HST} detectors and temporal variation of the PSFs. The stellar positions are then corrected for geometric distortion by means of the solutions provided for ACS/HRC and ACS/WFC by \citet{Anderson2004} and  \citet{Anderson2006}, respectively, and for WFC3/UVIS by \citet{Bellini2009}, and \citet{Bellini2011}. 
The final positional accuracy in an \textit{HST} image varies with the flux, but is around 0.01 pixels for faint well-measured stars ($\gtrsim 10,000$ or $\gtrsim 7,000$ counts in ACS/WFC and WFC3/UVIS, respectively), and less than 0.005 pixels just before saturation ($\sim 155,000$ or $\sim 105,000$ counts in ACS/WFC and WFC3/UVIS, respectively). Given the ACS/WFC and WFC3/UVIS pixel scales of 0.05 and 0.04 arcsec/pix, respectively, this translates into a typical positional accuracy ranging from 0.5 to 0.25 mas for the ACS and 0.4 to 0.2 mas for the WFC3.

\subsection{Combining {HST} and \textit{Gaia}: Reference Frame}\label{sec:reference_frame}

To combine \textit{HST} and \Gaia{} observations, \gaiahub{} needs to establish a common reference frame. This is done via 6-parameter linear fitting between the \textit{HST} and \Gaia{} positions at the different epochs. The solution of this fit includes zero point shift, rotation, scaling, and skew, and is used to transform the \textit{HST}-based positions onto the reference frame of \Gaia{} $(r_{\rm HST})$. The relative PMs are then computed as the difference between the \Gaia{} $(r_{\rm Gaia})$ and the transformed \textit{HST} positions, divided by the temporal baseline:

\begin{equation}\label{eq:pm}
    \mu = (r_{\rm Gaia} - r_{\rm HST})/\Delta T,
\end{equation}

where $\Delta T$ is the time baseline between both observations. Note that this method assign PMs even to the faint stars that have only positions but not PMs in the \Gaia{} catalog.

The solution of the 6-parameter linear fitting is the one that minimizes the differences in the relative positions of the stars between epochs. This is equivalent to assuming that the average difference in positions is zero, and that any relative difference in positions for a given star is the result of its peculiar motion with respect to the entire population. This provides good results when the motions of the stars are random and there are a statistically large number of them, which is the case for most targets in \gaiahub's expected uses. However, this minimization could potentially introduce spurious terms in the solution in certain cases where these conditions are not met. For example, the differential motion of MW stars along a specific direction could introduce rotation terms in the solution to the fitting that would result in a false counter rotation for our stars of interest.

To avoid this problem, the positions of the stars in the second epoch can be shifted back to the positions they were in the first epoch before setting up the reference frame. \gaiahub{} can perform this operation using the PMs of the stars derived in a first iteration and then refine the reference frame in subsequent iterations until convergence. This method provides good results when the PM uncertainties are small and therefore the projection of the positions from the latest epoch to the previous ones is accurate. However, large PMs uncertainties can cause the method to diverge, yielding worse results than those obtained without shifting back the position of the stars.

Another way to solve this problem is to only use co-moving stars to set up the reference frame between \textit{HST} and Gaia. To do this, \gaiahub{} can automatically select co-moving stars and use them to set up the reference frame (see Appendix~\ref{Apx:Membership_selection}). Such selection is made based on their PMs, which are then refined in subsequent iterations until convergence. Despite using fewer stars in the fitting, this option usually yields better results in fields with a significant amount of contaminant stars, as member stars show coherent motion in the sky with smaller velocity dispersion than the entire set of stars. Therefore it is important to consider the conditions of the target when setting up a reference frame and when deciding which of the above methods is most suitable. 

After computing the relative PMs, absolute ones are computed by just adding the average difference between the \Gaia{} PMs and the relative \textit{HST}- \Gaia{} PMs for all stars in the field with both measurements.

\subsection{Expected Nominal Accuracy}\label{sec:formal_errors}

\begin{figure}
\begin{center}
\includegraphics[width=\linewidth]{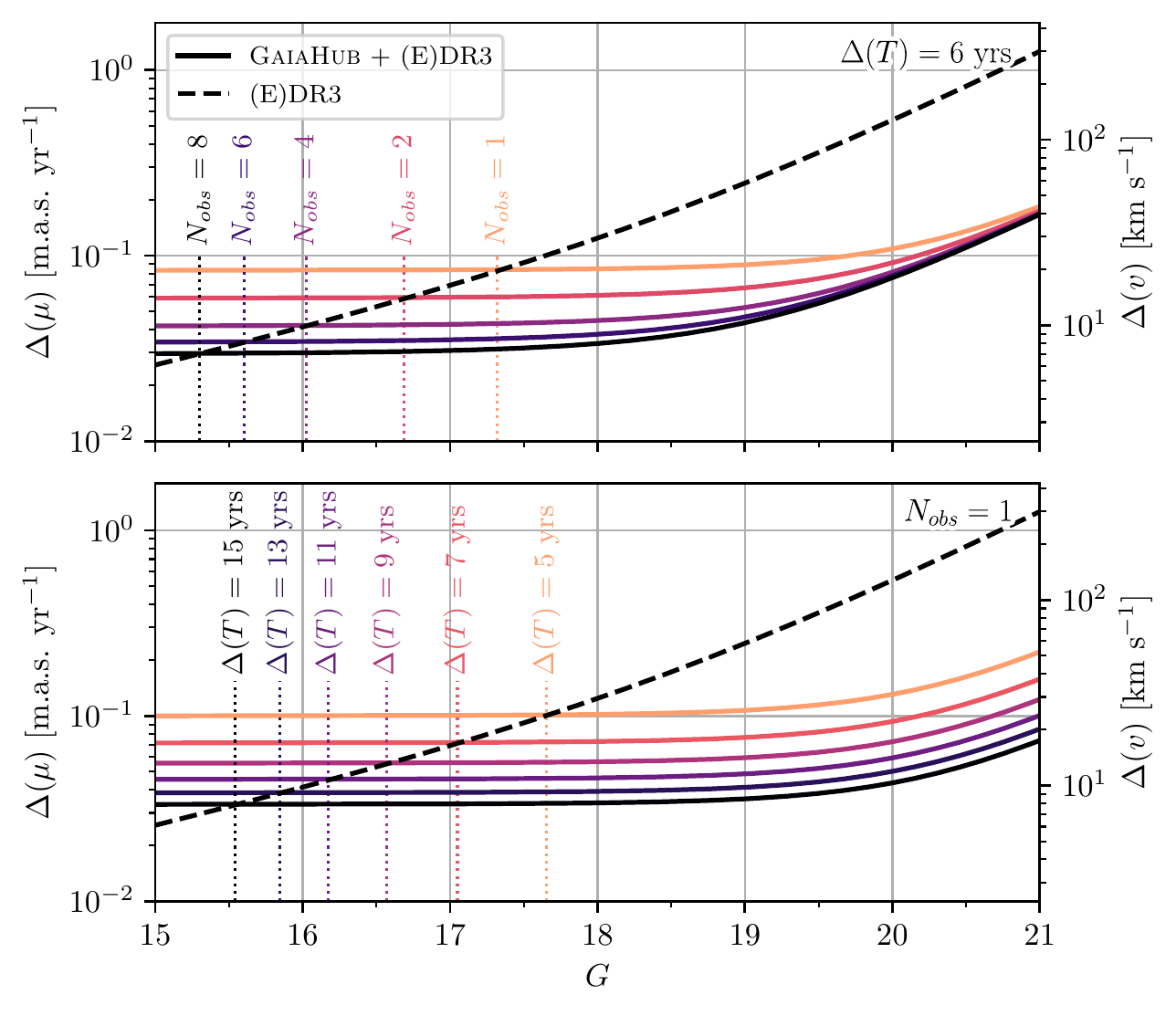}
\caption{The expected nominal PMs uncertainties of \gaiahub{} in different cases when one or more non-saturated \textit{HST} images has been combined with \Gaia{} EDR3. In all cases, we assume a constant $\Delta r_{\rm HST, i} = 0.5 \mas$ per \textit{HST} exposure (typical positional uncertainty for a faint, well measured star. See Section~\ref{sec:HST_astrometry}), while $\Delta r_{\rm Gaia}$ varies with $G$ assuming all-sky average values computed with \textsc{Pygaia}. The top panel shows the impact of using one or more dithered \textit{HST} images, $N_{obs}$, taken at the same epoch, June 2011. The bottom panel shows the impact of $\Delta T$ when using just one \textit{HST} image taken on different years (the typical baseline found in the data is $\sim11$ years). We use the last EDR3 data collection date, May 28th 2017, as epoch for the \Gaia{} positions. In both panels, nominal errors of EDR3 are shown by a black dashed curve. The intersection of the \gaiahub{} nominal errors with the \Gaia{} EDR3 nominal error is shown by vertical dashed lines.}
\label{fig:expected_error_dr3}
\end{center}
\end{figure}

\begin{figure}[ht!]
\begin{center}
\includegraphics[width=\linewidth]{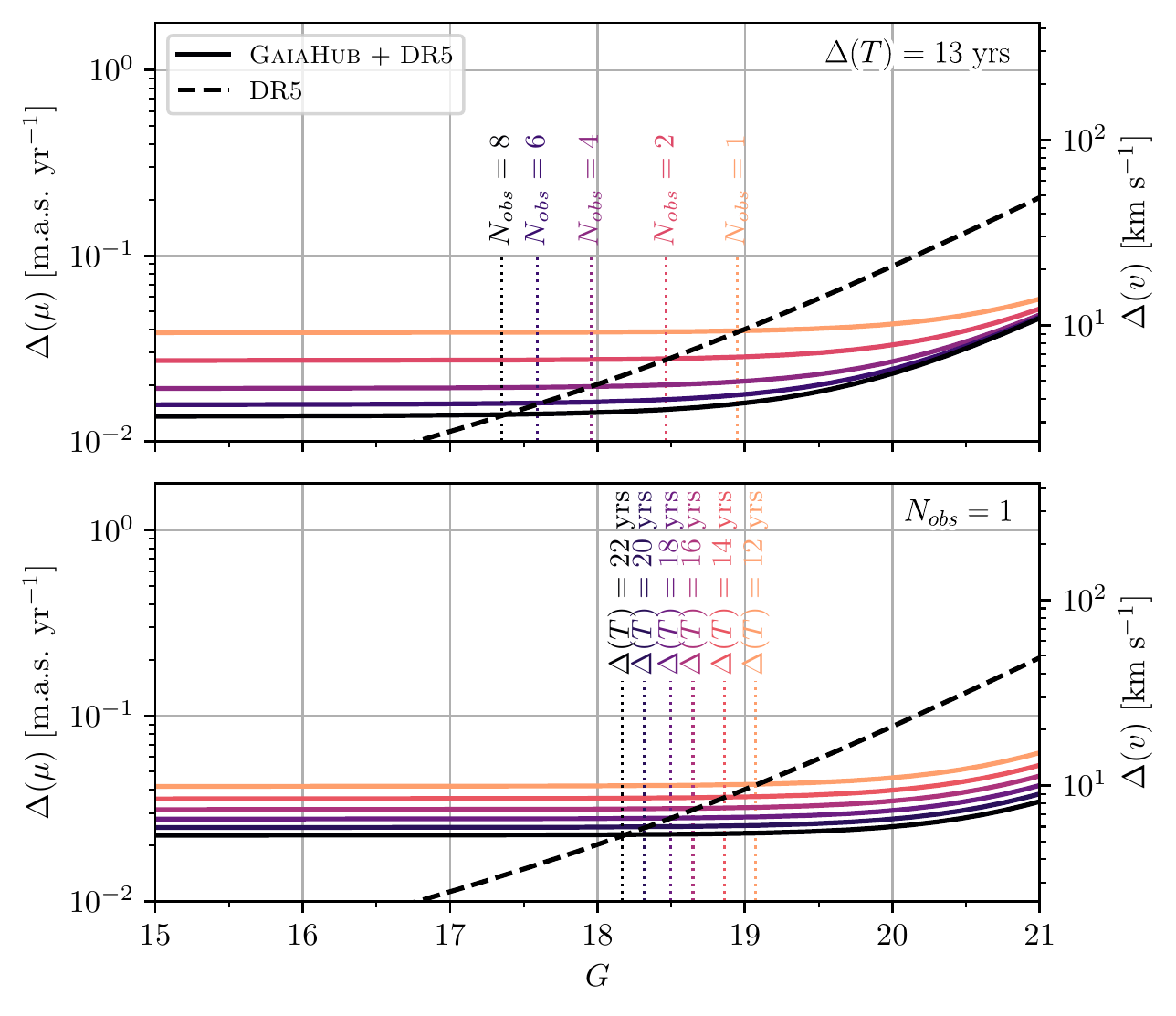}
\caption{Same as Figure~\ref{fig:expected_error_dr3}, but for \Gaia{} DR5 based on 10 years of data acquisition.}
\label{fig:expected_error_dr5}
\end{center}
\end{figure}

The random uncertainties on the PMs are the sum in quadrature between the \Gaia{} and \textit{HST} positional errors, divided by the temporal baseline,

\begin{equation}\label{eq:pm_uncertainty}
    \Delta\mu = \left[\Delta r_{\rm Gaia}^2 + \Delta r_{\rm HST}^2)\right]^{1/2}/\Delta T.
\end{equation}

Here, $\Delta r_{\rm HST}$ includes the error of the 6-parameter transformation, which decreases by a factor $\sqrt{N_\star}$, being $N_\star$ the number of stars used in the fit. 
The contribution of the transformation error has thus a negligible impact on the final result when a sufficiently large number of stars are used. For example, when 100 stars are used in the transformation, the contribution to the error of the \textit{HST} positions in the \Gaia{} reference frame is 10 times smaller than the original \textit{HST} positional error\footnote{By default, \gaiahub{} will require at least 10 stars in common between \HST{} and \Gaia{} to perform the fitting. \gaiahub{} options allow for an even lower number of stars, although it is not recommended. In any case, a minimum of 3 stars is required to obtain the 6-parameters transformation}. If $N_{obs}$ well-dithered \textit{HST} exposures are available, then the positional accuracy in the \textit{HST} data will 
decrease as $\Delta r_{\rm HST} = \sigma(r_{\rm HST})/\sqrt(N_{obs})$ where $\sigma(r_{\rm HST})$, is the root mean square (RMS) scatter of the stellar positions between the $N_{obs}$ exposures. 

Both the $\Delta r_{\rm Gaia}$ and $\Delta r_{\rm HST}$ positional uncertainties are potentially complicated depending on several factors. For example, \textit{HST} positional accuracy depends on the number of counts, hence on the filter and the integration time, but also on the position of the star in the CCD. \Gaia{} positional accuracy is also complex, and depends on the magnitude of the star, its color, and position in the sky among several other factors. However, the all-sky average of \Gaia{} uncertainties is well known, and if we consider only non-saturated stars in the \textit{HST} image, we can assume a constant $\Delta r_{\rm HST, i} = 0.5 \mas$ per exposure as a typical positional accuracy for \textit{HST}. Using these two quantities we can make some predictions about the expected accuracy of \gaiahub.

Figure~\ref{fig:expected_error_dr3} shows the expected PMs and velocity accuracies of \gaiahub{} for a bright, non-saturated star located at 50 kpc from the Sun in different cases where one or more \textit{HST} images are combined with \Gaia{} EDR3. In all cases, we assume a constant $\Delta r_{\rm HST, i} = 0.5 \mas$ per \textit{HST} exposure, while $\Delta r_{\rm Gaia}$ varies with $G$ assuming all-sky average values provided by Pygaia\footnote{\url{https://pypi.org/project/PyGaia/}}.
Adding more \textit{HST} images improves the final accuracy, so does increasing $\Delta T$. These two factors will have the largest impact on the final accuracy of \gaiahub{} results.

Because of the constant uncertainty for \textit{HST} positions, $\Delta r_{\rm Gaia}$ dominates at faint magnitudes and $\Delta r_{\rm HST}$ at brighter ones. The effect of this can be seen when comparing results for 4 \textit{HST} images taken in 2009 with those using a single \textit{HST} image from 2002. The former will provide better accuracy at magnitudes brighter than $G\sim18$ when combined with EDR3, but not at fainter magnitudes.

The results from these models show that \gaiahub{} will consistently provide better results than \Gaia{} alone for stars fainter than $G = 18$ when non-saturated \textit{HST} images are used. With future \Gaia{} data releases, the intersection between the two curves will move to fainter magnitudes as the ratio $\Delta T_{\rm Gaia + HST}/\Delta T_{\rm Gaia}$ decreases, but \gaiahub{} will always perform better than \Gaia{} alone for stars fainter than a given magnitude. Figure~\ref{fig:expected_error_dr5} shows the same accuracy model we used to describe the expected accuracy with EDR3 but for a future fifth \Gaia{} data release (DR5), based on 10 years of data acquisition\footnote{\url{https://www.cosmos.esa.int/web/gaia/release}}.

In practice, several \textit{HST} observations may exist for a given object, which can result in data from different epochs and quality. For example, images with different filters or integration times will result in data with different positional accuracy. In these cases, or even when only one \textit{HST} image is available, \gaiahub{} uses the quality of PSF fit to assign positional uncertainties to individual stars in individual images. The PM of a source and its uncertainty is then measured for each HST image individually with Equations~\ref{eq:pm} and~\ref{eq:pm_uncertainty}, which also take into account different time baselines. The final PM for a given star is finally computed as the error-weighted mean of the PMs corresponding to each \textit{HST} image. The corresponding final uncertainty is computed analytically propagating all known uncertainties and taking into account that measurements using several \HST{} epochs are correlated; i.e., all use \Gaia{} as second epoch.

\subsection{\gaiahub{} Execution}

\gaiahub{} is publicly available on Zenodo\footnote{\url{https://doi.org/10.5281/zenodo.6467326}} \citep{GaiaHub_zenodo} and Github\footnote{\url{https://github.com/AndresdPM/GaiaHub}} and is free to use and modify. The instructions for its installation and execution are included in the same repository. The code is written in Python and Fortran, and has been conceived as an automatic processing pipeline. Its most important components include the

\begin{itemize}
    \item download of the data,
    \item matching of the different observation epochs,
    \item membership selection of stars,
    \item and the computation of the PMs.
\end{itemize}

\gaiahub{} can be called from a terminal or from a script using the following syntax:

\begin{verbatim}
$ gaiahub [OPTIONS...]
\end{verbatim}.

In Appendix \ref{Apx:GaiaHub_Execution} we provide a general overview of \gaiahub{} and the options related to the specific parts of its execution.

\section{Results}\label{sec:Results}

We apply our approach here to samples of MW GCs and classical dwarf satellite galaxies. The goal is to illustrate the capabilities and demonstrate the accuracy of our approach. Scientific exploitation of the results for these and other objects will be presented in future follow-up papers.

\subsection{Globular Clusters}

GCs are dense systems with little to no dark matter, and are amongst the oldest stellar systems in the Universe. It is the combination of their density and longevity that makes them interesting from a dynamical standpoint. \textit{HST} has already successfully measured internal PM kinematics in GCs. However, this is only possible where there are multiple epochs of \textit{HST} data, well-separated temporally so as to give a long baseline for PM measurements \citep[see for example][]{Bellini2014}, and such data does not exist for many GCs.

\Gaia{} has also been used to study the internal kinematics of GCs, including their rotation on the plane of the sky \citep{Bianchini2018} and their velocity dispersion \citep{Vasiliev2021}. However, given its still relatively large astrometric errors, the results have been limited to nearby GCs ($\dsun \lesssim 25 \kpc$) and have only used relatively bright stars ($G \lesssim 20$).

The results from \gaiahub{} can complement these observations, enabling the measurement of precise PMs for stars that are too faint for \Gaia{} alone. Because of the improved accuracy, the technique is relevant for measuring velocity dispersions in many globular clusters, especially those located at large distances from the Sun. As a demonstration, from the \citet{Harris1996} catalog (2010 edition) we selected all the GCs located at heliocentric distances larger than $\dsun = 25 \kpc$, and those with only one \textit{HST} epoch available.
We also considered Omega Centauri, in order to compare our results with those obtained from \textit{HST}-only PMs. This resulted in a list of 40 GCs for which we ran \gaiahub{} to derive their internal PMs.

We ran \gaiahub{} on all of them using the default options and the automatic membership selection, which uses only member stars to perform the alignment between epochs\footnote{This is accomplished by running \gaiahub{} with the \mbox{\tt --use\_members} flag, as described in the Appendix~\ref{Apx:GaiaHub_Execution}}. We selected the oldest \textit{HST} observations in cases where there is more than one \textit{HST} epoch for a given field. This reduces the number of images that \gaiahub{} has to process and does not impact the final accuracy by much. MW contaminant stars can be a problem in those clusters located at small Galactic latitudes. To avoid selecting them as members, we forced \gaiahub{} to clip the selection at $2.5\sigma$ instead of the default $3\sigma$\footnote{Using the \mbox{\tt --clipping\_prob\_pm 2.5} flag.}. In cases where MW stars dominate the field and the automatic membership selection does not converge or selects MW stars as members; i.e. Djorg 1, NGC\ 6642, NGC\ 6558, and Lynga 7; we manually selected the approximate area in the PMs vector-point diagram (VPD) where the member stars are located prior to the automatic membership selection\footnote{Using \mbox{\tt --preselect\_pm} flag.}. Lastly, in the case of NGC\ 2419 we decided not to use images in the F555W filter, as they produced results that were not compatible with those obtained with the rest of the filters.

\gaiahub{} provided clearly better results than \Gaia{} alone in all clusters except for AM4, where the scarce number of member stars in common between \textit{HST} and \Gaia{} (11 stars), together with the relatively short increment in the time baseline (5.8 years compared to 2.8 years for EDR3 alone) did not yield a visibly lower dispersion of the PMs in the VPD. The clusters NGC\ 6624 and NGC\ 6401 did not converge to a solution. The \textit{HST} fields for both clusters have an extremely high density of stars, both members and MW contaminants, and we believe this causes \gaiahub{} to incorrectly pair stars, producing spurious results. Hence, we decided to remove these three clusters from our final list, which ended up containing 37 clusters.

\subsubsection{Individual Examples}

\begin{figure}
\begin{center}
\includegraphics[width=\linewidth]{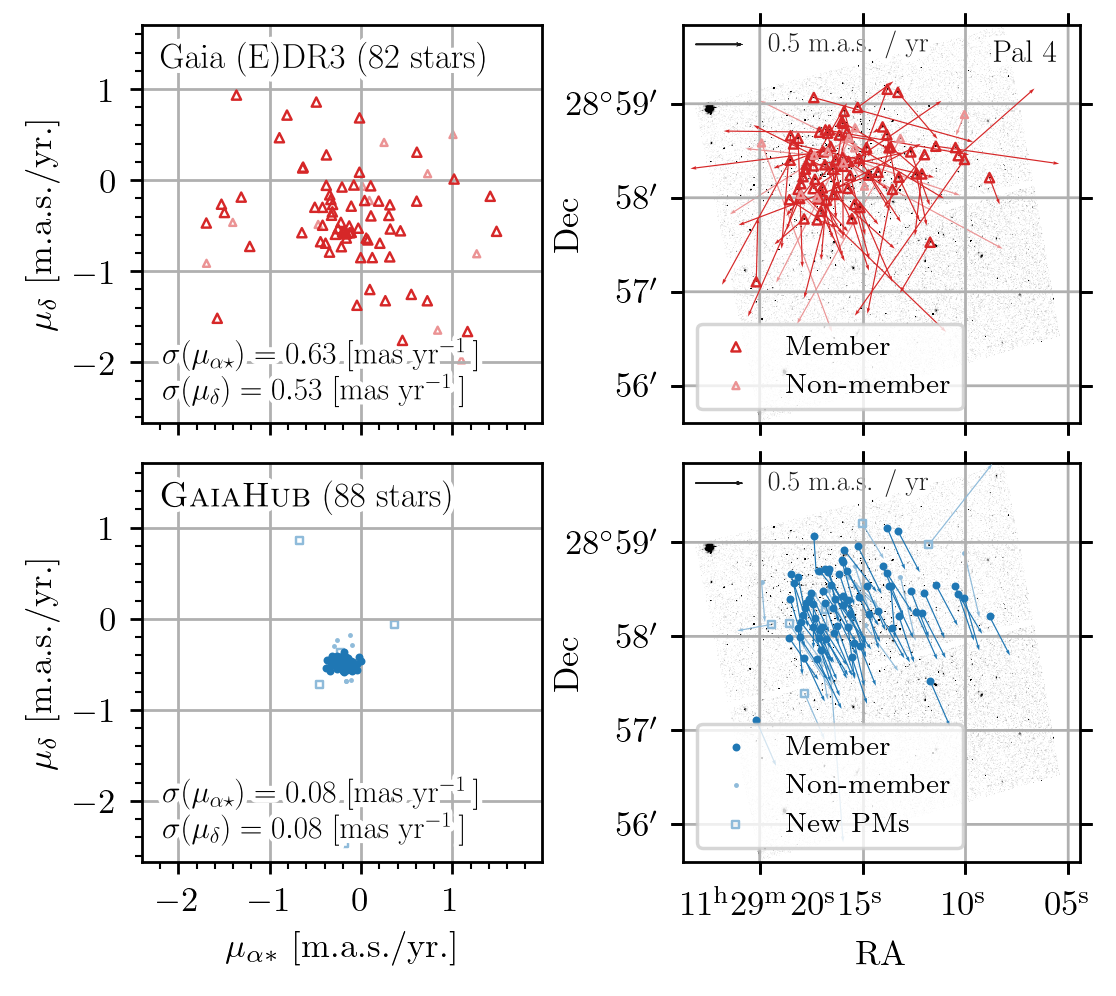}
\caption{Comparison between the results obtained using Gaia and GaiaHub for Palomar 4 ($\dsun = 108.7\kpc$, $\Delta T = 11.2$ years). The VPDs are shown in the left column and the PMs projected on the sky in the right column. Results from \Gaia{} are represented in the top row by red symbols, new results from \gaiahub{} are shown by blue symbols in the bottom row. New PMs, not present in the \Gaia{} EDR3 catalog are shown by open blue squares. Member stars, automatically selected by \gaiahub{} based on position in the VPD, are represented by large, darker markers. The larger dispersion observed in the \Gaia{} VPD and PMs is a consequence of its larger uncertainties.}
\label{fig:Comparative_Gaia}
\end{center}
\end{figure}

Here we present four examples of the quality of the results obtained with \gaiahub{} in comparison with those of \Gaia{} EDR3 alone. In Figure~\ref{fig:Comparative_Gaia} we show the impact that the precision of the PMs measurements has on the study of the internal kinematics of a stellar system by comparing \Gaia's VPD and the vectorial representation of its PMs in the observed field, with those obtained using \gaiahub{} in GC Palomar 4. Larger uncertainties in \Gaia's PMs are reflected in the perceived motion of the stars, and could lead to artificially large velocity dispersion measurements in stellar systems. Below, we show a more detailed comparison for three GCs ordered by their heliocentric distance.

\paragraph{NGC\ 6535}

\begin{figure*}[ht!]
\begin{center}
\includegraphics[width=\linewidth]{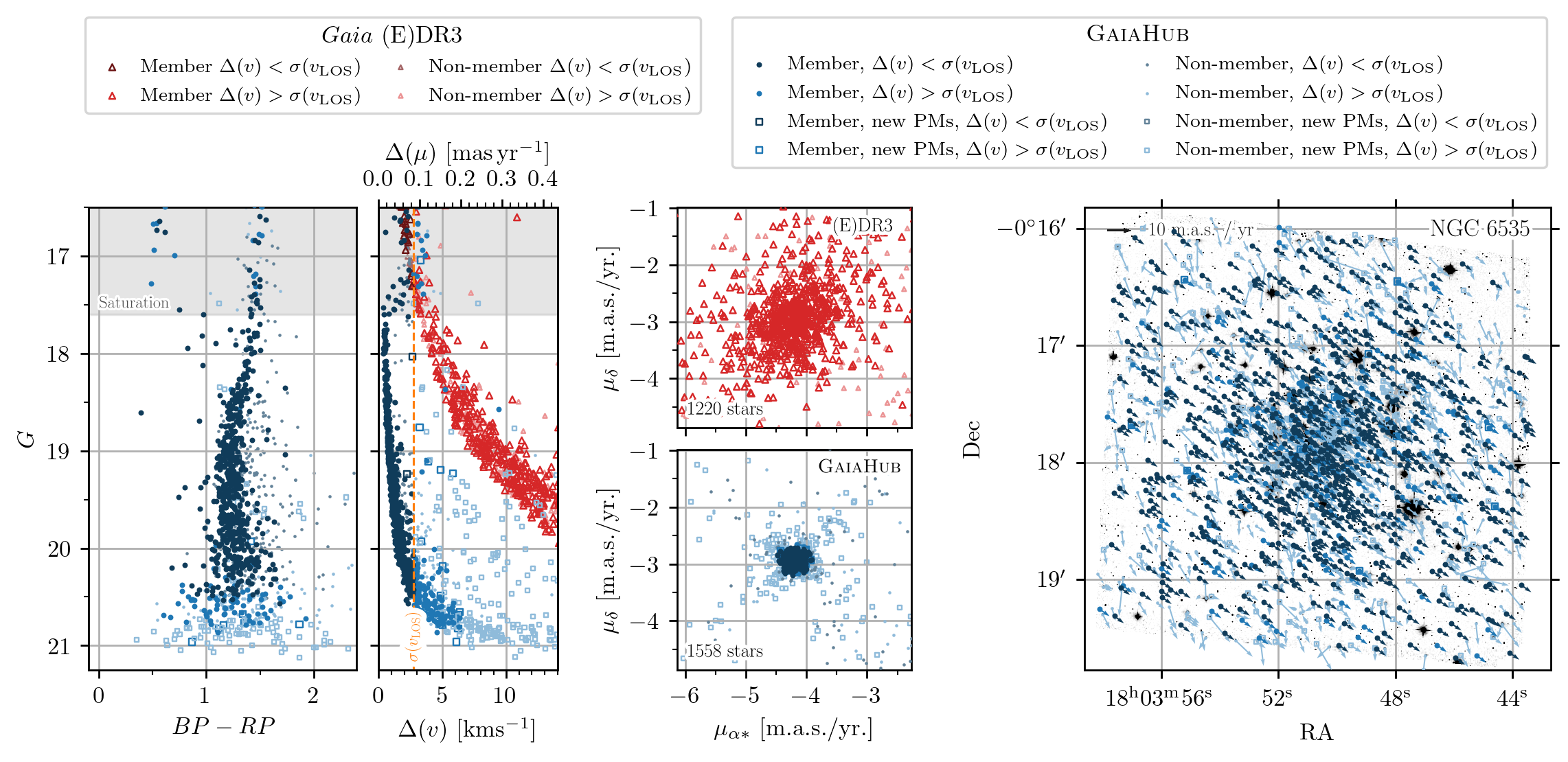}
\caption{Summary of the results for NGC\ 6535 using \gaiahub{} and \Gaia{} alone. From left to right the panels show the color-magnitude diagram (CMD), the PM uncertainties as a function of $G$ mag, the VPD for \Gaia{} EDR3 (top) and \gaiahub{} (bottom), and the footprint of the analyzed area showing the PMs as vectors. Results from \Gaia{} are represented in the second and third columns by open red triangles. New results from \gaiahub{} are shown by blue symbols. New PMs, not present in the \Gaia{} EDR3 catalog are shown by open blue squares. Member stars, automatically selected by \gaiahub{} based on position in the VPD, are represented by large, darker markers. Stars whose PM uncertainty is below the central $\svlos$ value of NGC 6535, marked by an orange vertical dashed line in the second panel ($2.4\pm0.5 \kms$, see Table~\ref{tab:summary_GCs}), are shown in dark blue. Non member stars are represented by light red (\textit{Gaia}) and light blue (\gaiahub). The gray shaded area in two leftmost panels indicates the region of saturation for the deepest HST images. In the VPD (third column of panels), the dark red open triangles from the top panel (\textit{Gaia}) compare to the dark blue dots from the bottom panel (\gaiahub). Notice that newly measured PMs that are not in the \Gaia{} catalog (light-blue open squares) are not present in the \textit{Gaia}'s VPD. The cross-like distribution observed in the VPD for very faint stars is likely caused by \Gaia's positional errors due to the observatory's scanning law \citep[see for example][]{Bianchini2018}. For clarity, the rightmost panel shows only measurements using \gaiahub.}
\label{fig:NGC6535_vpd}
\end{center}
\end{figure*}

Located at $\dsun = 6.8 \kpc$, NGC\ 6535 provides a good example of the improvements in the PMs of faint stars in relatively nearby systems. The first \textit{HST} epoch was taken in late March 2006, providing a total time baseline of 11.2 years. A comparison between the results from \Gaia{} and \gaiahub{} is shown in Figure~\ref{fig:NGC6535_vpd}. \gaiahub{} clearly outperforms \Gaia{} for magnitudes $G \gtrsim 17.25$, which in NGC\ 6535 includes almost all the observed stars below the horizontal branch (HB). The uncertainties are much smaller than those of \Gaia{} alone, keeping values under the central line-of-sight velocity dispersion, $\svlos = 2.4\pm0.4 \kms$, up to $G\sim 20.5$, well below the main sequence turn-off (MSTo). For comparison, the mean PM uncertainty for \Gaia{} at $G\sim 20.5$ is $\sim 40 \kms$. The improved precision can be appreciated in the VPD, with \gaiahub's PMs being much more concentrated. \gaiahub{} also derived PMs for 338 stars that have no PMs in the \Gaia{} catalog. Despite having on average larger uncertainties than the rest of the \gaiahub's measurements ($5 \kms \lesssim \Delta(v) \lesssim 15 \kms$) these newly measured PMs still have smaller uncertainties than most \Gaia{} stars fainter than $G\sim19.5$.

\paragraph{NGC 5053}

\begin{figure*}[ht!]
\vspace{-10 mm}
\begin{center}
\includegraphics[width=\linewidth]{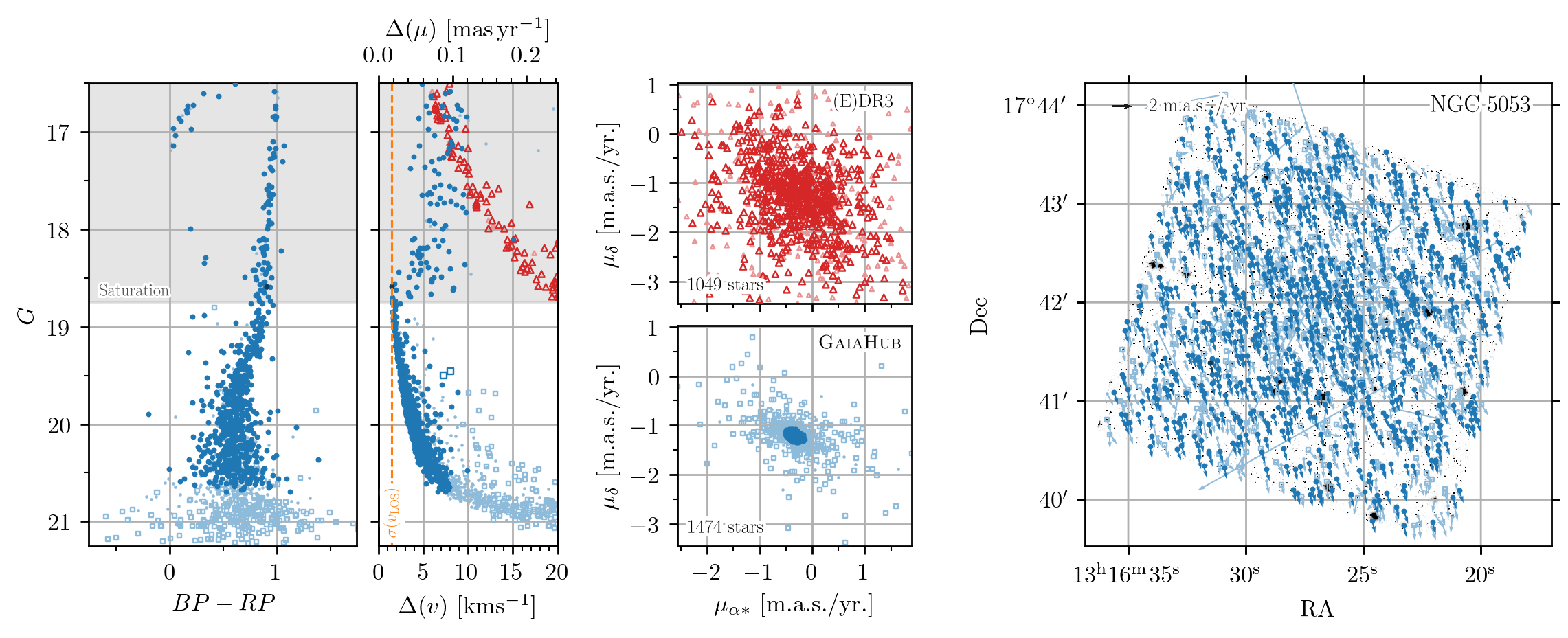}
\vspace{-3 mm}
\caption{Summary of the results for NGC\ 5053 GC. Markers and colors coincide with those of Figure~\ref{fig:NGC6535_vpd}.}
\label{fig:NGC5053_vpd}
\end{center}
\end{figure*}

In Figure~\ref{fig:NGC5053_vpd}, we show a summary of the results obtained for NGC\ 5053. This is a relatively metal-poor cluster ($[Fe/H] = -2.27$) with very low velocity dispersion, $\svlos = 1.4\pm0.2$. Because of this and the relatively large distance to the cluster, only one star has PMs uncertainties below the $\svlos$ value. However, \gaiahub{} provides results far better than those of \Gaia{} alone and allow to derive new PMs for 425 stars.

\paragraph{Pal 2}

\begin{figure*}[ht!]
\begin{center}
\includegraphics[width=\linewidth]{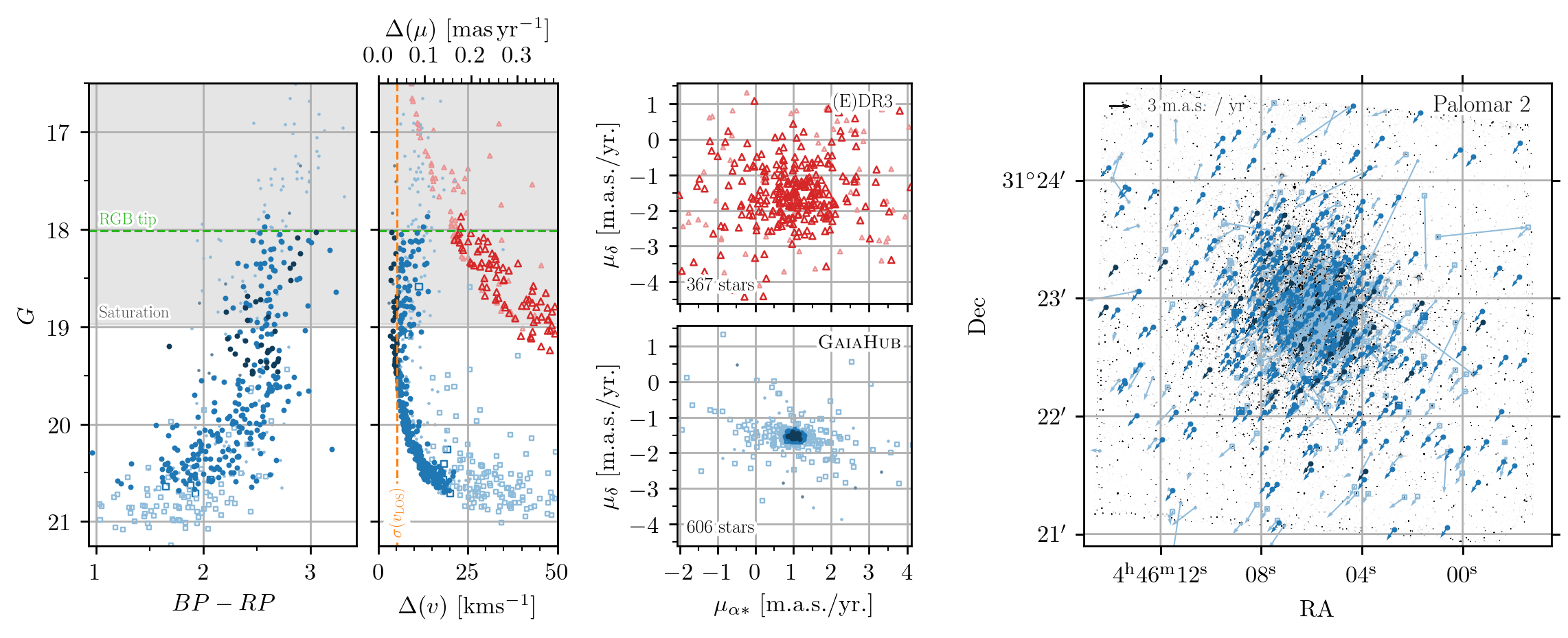}
\caption{Summary of the results for Pal 2 GC. Markers and colors coincide with those of Figure~\ref{fig:NGC6535_vpd}. The RGB tip magnitude is shown by a green dashed horizontal line in the two leftmost panels. The central value of the central $\svlos = 5.1 \kms$ is taken from \citet{Baumgardt2018} (3rd version of the catalog, May 2021).}
\label{fig:Pal 2_vpd}
\end{center}
\end{figure*}

Pal 2 is located at 27.2 kpc from the Sun, which makes it a good target for \gaiahub. Figure~\ref{fig:Pal 2_vpd} summarizes the results for \gaiahub. Despite the relatively large distance, \gaiahub{} allows us to derive PMs with uncertainties below $\svlos$ values for many of its stars, as well as to derive new PMs for 239 stars. 

\paragraph{NGC\ 2419}

\begin{figure*}[ht!]
\begin{center}
\includegraphics[width=\linewidth]{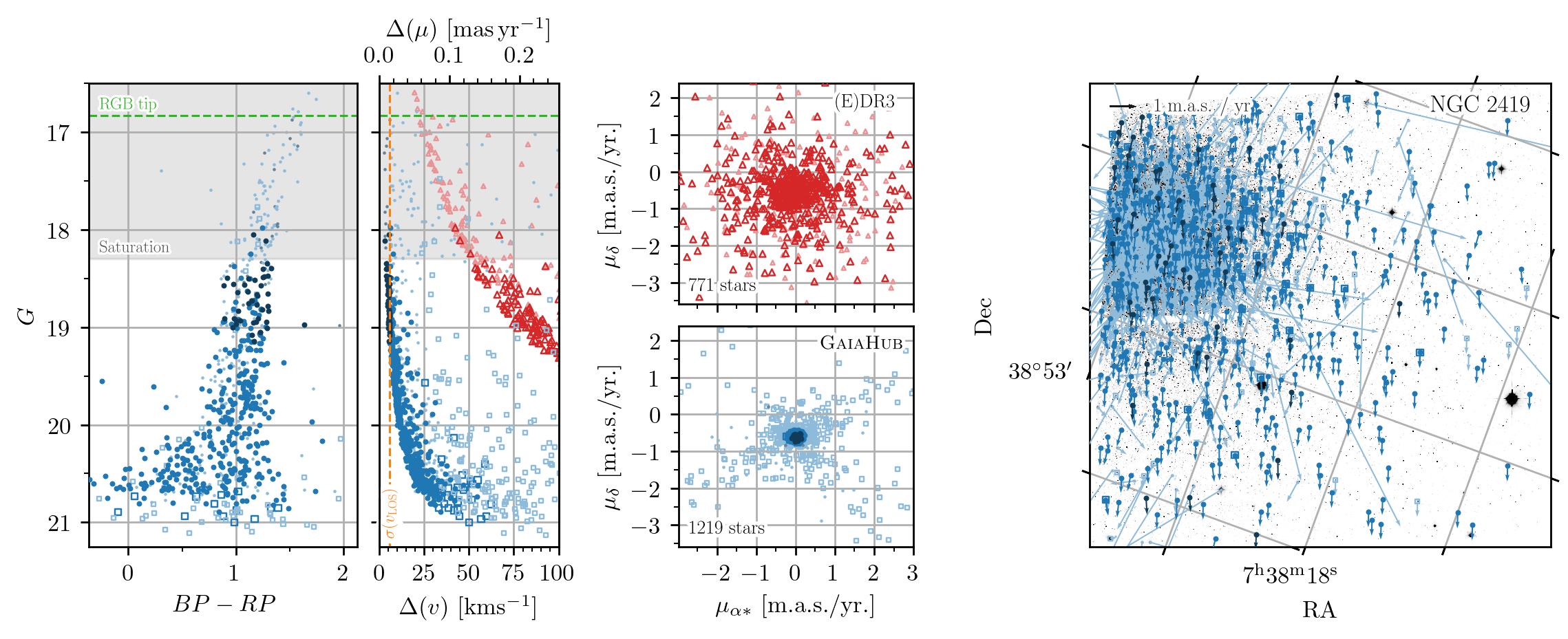}
\caption{Summary of the results for NGC\ 2419. Markers and colors coincide with those of Figure~\ref{fig:NGC6535_vpd}.}
\label{fig:NGC2419_vpd}
\end{center}
\end{figure*}

The results for NGC\ 2419 are shown in Figure~\ref{fig:NGC2419_vpd}. Despite the large distance to NGC\ 2419, $\dsun = 82.6 \kpc$, \gaiahub{} reaches accuracies at the level of its line-of-sight velocity dispersion, $\svlos = 4.0\pm0.6 \kms$, for some of its stars, while showing average accuracies of $\sim 11 \kms$ at $G=20$. At the same magnitude, \Gaia{} alone shows uncertainties of $\Delta(v) \sim 180 \kms$. The smallest uncertainty reached by \Gaia{} EDR3, $\Delta(v) \sim 20 \kms$, is for the brightest stars at the tip of the red giant branch (TRGB) at $G \sim 16.75$. Future \Gaia{} data releases will allow \gaiahub{} to measure hundreds of NGC\ 2419 stars below $\svlos$ levels.

\subsubsection{Dispersion, Anisotropy, and Ellipticity}\label{sec:GCs_anisotropy}

For the 37 clusters in our final list, we used the PMs derived with \gaiahub{} to calculate the mean internal velocity dispersion along the radial, $\spmr$, and tangential, $\spmt$, directions with respect to the cluster's center. To do so we use a maximum likelihood approach \citep[as described in Section 3.1 of][]{Watkins2015} that properly accounts for inflation of the observed PM dispersions by observational error. We then computed the velocity dispersion values in each direction as $\sigma(v) = 4.7404 \times \dsun \times \sigma(\mu)$ and the sky-projected anisotropy, $\beta_{\rm sky} = 1 - \svt^2/\svr^2$. All sources of random errors were propagated following a Monte Carlo scheme in the case of the velocity dispersion, and analitically for the later computation of $\beta_{\rm sky}$\footnote{Uncertainties for $\beta$ obtained through our error propagation scheme are not constrained in any way, which may results in uncertainties that are compatible with nonphysical values, i.e. $\beta_{\rm sky} > 1$.}, assuming a 10\% error in the distance to the clusters. Our results are listed in Table~\ref{tab:summary_GCs}.
 
\begin{deluxetable*}{ccccccccccc}
\tabletypesize{\scriptsize}
\tablenum{1}
\tablecaption{PM uncertainties and dispersions for GCs using \gaiahub{} and \Gaia{} EDR3.}
\label{tab:summary_GCs}
\tablewidth{0pt}
\tabcolsep=0.11cm
\tablehead{GC & $\dsun$ & $\Delta T$ & $\Delta(v)_\mathrm{Gaia}$ & $\Delta(v)_\mathrm{\gaiahub}$ & $\spmr$ & $\spmt$ & $\svlos$ &  $\svr$ & $\svt$ & $\beta_{\rm sky}$ \\
                  &   (kpc) & yr & $(\kms)$ & $(\kms)$ & $(\masyr)$  & $(\masyr)$ & $(\kms)$  & $(\kms)$  & $(\kms)$  &   }
\decimalcolnumbers
\startdata
NGC6397 & 2.3 & 12.2 & 7.9 & 1.0 & $0.352 \pm 0.005$ & $0.334 \pm 0.005$ & $4.50 \pm 0.20$ & $3.8 \pm 0.4$ & $3.6 \pm 0.4$ & $0.10 \pm 0.04$ \\
NGC6656 & 3.2 & 12.8 & 17.8 & 7.7 & $0.408 \pm 0.008$ & $0.436 \pm 0.009$ & $7.80 \pm 0.30$ & $6.2 \pm 0.6$ & $6.6 \pm 0.7$ & $-0.14 \pm 0.06$ \\
NGC6752 & 4.0 & 11.6 & 12.9 & 1.2 & $0.256 \pm 0.005$ & $0.227 \pm 0.005$ & $4.9 \pm 0.4$ & $4.9 \pm 0.5$ & $4.3 \pm 0.4$ & $0.21 \pm 0.05$ \\
NGC5139 & 5.2 & 11.4 & 15.2 & 6.5 & $0.370 \pm 0.012$ & $0.329 \pm 0.011$ & $16.80 \pm 0.30$ & $9.1 \pm 1.0$ & $8.1 \pm 0.9$ & $0.21 \pm 0.07$ \\
NGC6535 & 6.8 & 11.2 & 21.2 & 1.9 & $0.0530 \pm 0.0024$ & $0.0562 \pm 0.0025$ & $2.4 \pm 0.5$ & $1.71 \pm 0.19$ & $1.81 \pm 0.20$ & $-0.13 \pm 0.14$ \\
Terzan5 & 6.9 & 13.7 & 27.9 & 3.8 & $0.385 \pm 0.008$ & $0.393 \pm 0.008$ &  & $12.6 \pm 1.3$ & $12.8 \pm 1.3$ & $-0.04 \pm 0.06$ \\
NGC6558 & 7.4 & 13.7 & 43.4 & 16.7 & $0.076 \pm 0.006$ & $0.095 \pm 0.007$ & $3.1 \pm 0.9$ & $2.68 \pm 0.35$ & $3.3 \pm 0.4$ & $-0.53 \pm 0.34$ \\
NGC6325 & 7.8 & 7.1 & 42.6 & 8.4 & $0.105 \pm 0.008$ & $0.109 \pm 0.008$ & $5.9 \pm 1.3$ & $3.9 \pm 0.5$ & $4.0 \pm 0.5$ & $-0.10 \pm 0.23$ \\
NGC6333 & 7.9 & 11.0 & 43.0 & 10.5 & $0.1660 \pm 0.0033$ & $0.1661 \pm 0.0034$ &  & $6.2 \pm 0.6$ & $6.2 \pm 0.6$ & $-0.00 \pm 0.06$ \\
Lynga7 & 8.0 & 11.1 & 34.7 & 2.6 & $0.0700 \pm 0.0025$ & $0.0715 \pm 0.0025$ &  & $2.66 \pm 0.28$ & $2.71 \pm 0.29$ & $-0.04 \pm 0.10$ \\
E3 & 8.1 & 11.1 & 18.5 & 1.7 & $0.0019 \pm 0.0021$ & $0.0020 \pm 0.0021$ &  & $0.07 \pm 0.08$ & $0.08 \pm 0.08$ & $-0.0 \pm 3.1$ \\
NGC6642 & 8.1 & 13.2 & 39.4 & 9.0 & $0.070 \pm 0.006$ & $0.063 \pm 0.006$ &  & $2.7 \pm 0.4$ & $2.41 \pm 0.35$ & $0.20 \pm 0.22$ \\
NGC6342 & 8.5 & 7.8 & 40.1 & 6.8 & $0.085 \pm 0.005$ & $0.082 \pm 0.005$ & $5.2 \pm 2.1$ & $3.4 \pm 0.4$ & $3.3 \pm 0.4$ & $0.06 \pm 0.16$ \\
NGC6355 & 9.2 & 7.8 & 53.9 & 12.4 & $0.127 \pm 0.007$ & $0.125 \pm 0.007$ &  & $5.5 \pm 0.6$ & $5.4 \pm 0.6$ & $0.03 \pm 0.16$ \\
NGC2808 & 9.6 & 10.8 & 36.5 & 4.0 & $0.1465 \pm 0.0029$ & $0.1390 \pm 0.0027$ & $13.4 \pm 1.2$ & $6.7 \pm 0.7$ & $6.3 \pm 0.6$ & $0.10 \pm 0.05$ \\
NGC6139 & 10.1 & 7.3 & 60.3 & 14.1 & $0.186 \pm 0.006$ & $0.198 \pm 0.006$ &  & $8.9 \pm 0.9$ & $9.5 \pm 1.0$ & $-0.14 \pm 0.10$ \\
NGC6256 & 10.3 & 7.8 & 52.7 & 7.1 & $0.153 \pm 0.007$ & $0.157 \pm 0.007$ & $6.6 \pm 2.6$ & $7.5 \pm 0.8$ & $7.7 \pm 0.8$ & $-0.05 \pm 0.13$ \\
NGC6517 & 10.6 & 7.1 & 50.5 & 10.6 & $0.172 \pm 0.006$ & $0.162 \pm 0.006$ &  & $8.6 \pm 0.9$ & $8.1 \pm 0.9$ & $0.11 \pm 0.10$ \\
NGC5946 & 10.6 & 7.8 & 49.8 & 10.4 & $0.120 \pm 0.005$ & $0.121 \pm 0.005$ & $4.0 \pm 2.9$ & $6.0 \pm 0.7$ & $6.1 \pm 0.7$ & $-0.01 \pm 0.12$ \\
NGC6380 & 10.9 & 7.2 & 49.6 & 13.2 & $0.183 \pm 0.006$ & $0.177 \pm 0.006$ &  & $9.4 \pm 1.0$ & $9.2 \pm 1.0$ & $0.06 \pm 0.09$ \\
Pal1 & 11.1 & 11.2 & 26.9 & 2.1 & $0.005 \pm 0.004$ & $0.005 \pm 0.004$ &  & $0.26 \pm 0.20$ & $0.25 \pm 0.19$ & $0.0 \pm 2.1$ \\
NGC6453 & 11.6 & 7.1 & 64.6 & 69.7 & $0.150 \pm 0.007$ & $0.175 \pm 0.008$ &  & $8.3 \pm 0.9$ & $9.6 \pm 1.1$ & $-0.36 \pm 0.19$ \\
Djorg1 & 13.7 & 13.3 & 63.8 & 5.7 & $0.103 \pm 0.008$ & $0.101 \pm 0.008$ &  & $6.7 \pm 0.8$ & $6.6 \pm 0.8$ & $0.03 \pm 0.21$ \\
NGC5466 & 16.0 & 11.1 & 46.8 & 3.5 & $0.0252 \pm 0.0016$ & $0.0245 \pm 0.0016$ & $1.70 \pm 0.20$ & $1.91 \pm 0.23$ & $1.86 \pm 0.22$ & $0.05 \pm 0.18$ \\
NGC5053 & 17.4 & 11.2 & 60.6 & 4.2 & $0.0175 \pm 0.0034$ & $0.009 \pm 0.004$ & $1.40 \pm 0.20$ & $1.45 \pm 0.31$ & $0.8 \pm 0.4$ & $0.72 \pm 0.28$ \\
NGC5024 & 17.9 & 11.2 & 87.2 & 7.9 & $0.0775 \pm 0.0025$ & $0.0739 \pm 0.0026$ & $4.4 \pm 0.9$ & $6.6 \pm 1.0$ & $6.3 \pm 1.0$ & $0.09 \pm 0.06$ \\
NGC6864 & 20.9 & 7.1 & 100.3 & 62.2 & $0.085 \pm 0.005$ & $0.084 \pm 0.005$ & $10.3 \pm 1.5$ & $8.4 \pm 1.0$ & $8.3 \pm 1.0$ & $0.02 \pm 0.16$ \\
Pal13 & 26.0 & 6.9 & 70.9 & 10.6 & $0.015 \pm 0.009$ & $0.015 \pm 0.010$ & $0.90 \pm 0.30$ & $1.8 \pm 1.2$ & $1.9 \pm 1.2$ & $-0.1 \pm 2.0$ \\
Terzan8 & 26.3 & 11.0 & 89.0 & 6.8 & $0.011 \pm 0.005$ & $0.015 \pm 0.005$ &  & $1.3 \pm 0.6$ & $1.8 \pm 0.7$ & $-0.8 \pm 2.1$ \\
NGC6715 & 26.5 & 11.0 & 120.3 & 15.2 & $0.0816 \pm 0.0020$ & $0.0839 \pm 0.0021$ & $10.50 \pm 0.30$ & $10.3 \pm 1.1$ & $10.5 \pm 1.1$ & $-0.06 \pm 0.07$ \\
Pal2 & 27.2 & 10.8 & 113.0 & 8.6 & $0.030 \pm 0.005$ & $0.029 \pm 0.005$ &  & $3.9 \pm 0.7$ & $3.7 \pm 0.7$ & $0.1 \pm 0.4$ \\
Arp2 & 28.6 & 11.1 & 90.1 & 8.1 & $0.013 \pm 0.006$ & $0.0025 \pm 0.0023$ &  & $1.7 \pm 0.8$ & $0.33 \pm 0.31$ & $0.96 \pm 0.08$ \\
NGC7006 & 41.2 & 7.6 & 123.9 & 19.2 & $0.048 \pm 0.006$ & $0.024 \pm 0.009$ &  & $9.4 \pm 1.5$ & $4.7 \pm 1.8$ & $0.74 \pm 0.19$ \\
Pal15 & 45.1 & 7.6 & 147.1 & 15.3 & $0.007 \pm 0.006$ & $0.006 \pm 0.006$ &  & $1.5 \pm 1.2$ & $1.4 \pm 1.2$ & $0.2 \pm 1.9$ \\
NGC2419 & 82.6 & 14.7 & 217.9 & 17.4 & $0.010 \pm 0.008$ & $0.015 \pm 0.009$ & $4.0 \pm 0.6$ & $3.9 \pm 3.2$ & $6 \pm 4$ & $-1.4 \pm 3.2$ \\
Pal4 & 108.7 & 11.2 & 291.4 & 29.9 & $0.011 \pm 0.006$ & $0.006 \pm 0.004$ &  & $5.5 \pm 3.0$ & $3.2 \pm 2.3$ & $0.6 \pm 0.6$ \\
\enddata
\tablecomments{Columns (2) and (8) are taken from \citet{Harris1996} (2010 version). Columns (4) and (5) are the median error in the velocity measured from PMs at $G=20$ with \Gaia{} EDR3 and \gaiahub, respectively. Columns (9) and (10) assume a 10\% error in the distance to the cluster (2). Column (11) is the sky-projected anisotropy.}
\end{deluxetable*}

\begin{figure}
\begin{center}
\includegraphics[width=\linewidth]{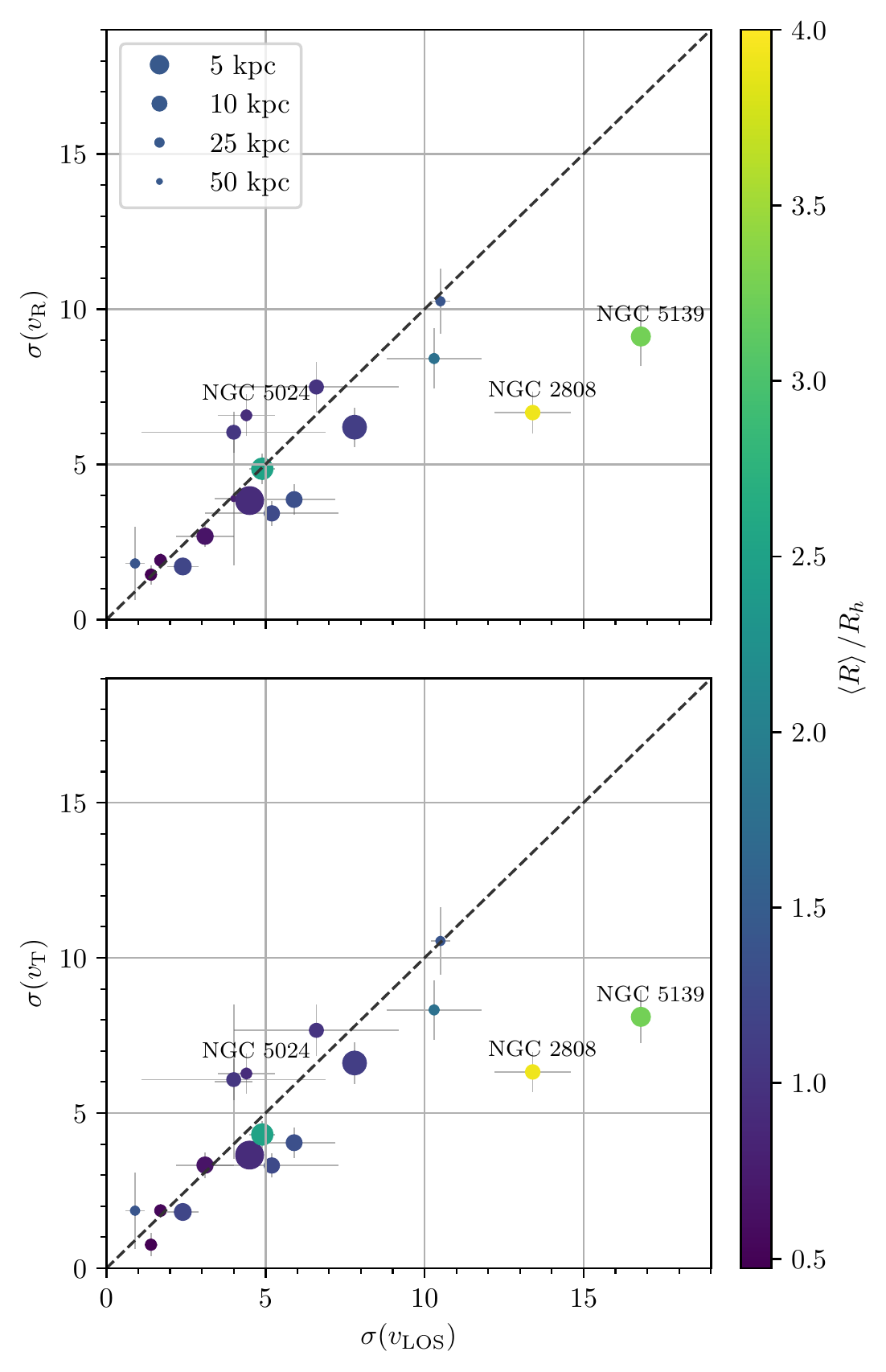}
\caption{Mean internal velocity dispersion along the radial, $\svr$, and tangential, $\svt$, directions with respect to the center, for the GC sample described in the text and listed in Table~\ref{tab:summary_GCs}. The color shade indicates the average distance of the analyzed data with respect to the center coordinates of the cluster. The size is inversely proportional to the Galactocentric distance to the cluster. The central velocity dispersion along the line-of-sight, $\svlos$, was taken from \citet{Harris1996} (2010 edition). The red dashed line shows a one-to-one relation. The data analyzed in NGC\ 2808 and NGC\ 5139 (Omega Cent) is located at 3.9 and 3.25 times their half-light radius, $R_h$, and therefore their $\svr$ and $\svt$ are expected to be smaller than $\svlos$. NGC\ 5024 shows values not compatible with $\svlos$ by $\sim2.5\sigma$, probably caused by systematic errors not included in its error bars.}
\label{fig:GCs_vrtlos}
\end{center}
\end{figure}

To assess the quality of the results, we compared the velocity dispersion along the three components in those clusters with available line-of-sight velocity dispersion, $\svlos$, in the \citet{Harris1996} catalog. This is shown in Figure~\ref{fig:GCs_vrtlos}. The dispersion along the three components is consistent, in general, except for NGC\ 2808, NGC\ 5139, and NGC\ 5024. Because $\svr$ and $\svt$ are mean values for all member stars within the \textit{HST} field, in general we expect the central values of $\svlos$ to be slightly larger. This is especially the case in clusters where the observed \textit{HST} field is located at a large radial distance from the center of the cluster, such NGC\ 2808 and NGC\ 5139 (Omega Centauri). In fact, also relatively low dispersion values have been reported in the same field in $\omega$ Centauri using \textit{HST}-only PMs \citep{Bellini2018}. That work separated the sample into multiple stellar populations, with the dominant one (MS-I) showing $\svr = 8.46 \pm 0.19 \kms$ and $\svt = 8.16 \pm 0.18 \kms$, both in very good agreement with the ones found using \gaiahub, i.e. $\svr = 9.1 \pm 1.0 \kms$ and $\svt = 8.1 \pm 0.9 \kms$.

It is worth noticing that the uncertainties listed in Table~\ref{tab:summary_GCs} do not include possible systematic errors. We expect their impact to be small in nearby clusters, but they could be non-negligible for distant clusters. A possible example is NGC\ 5024, which shows an unusually large $\svr$ and $\svt$; more than $\sim2.5\sigma$ larger than $\svlos$. Lower values in $\svr$ and $\svt$ can be obtained with a more restrictive membership selection. For example, clipping the selection at $2\sigma$ instead of the $2.5\sigma$ used for the rest of the clusters produces $\svr$ and $\svt$ values that are consistent with $\svlos$. In any case, it seems clear that NGC\ 5024 suffers from non-negligible systematic errors compared to its random uncertainties, and that this could also be the case for some other distant clusters. Moreover, because this is a demonstration paper for \gaiahub, we do not perform a thorough membership selection or error clipping prior to the $\spmr$ and $\spmt$ calculation, but use instead the automatic selection performed by \gaiahub. Hence, the results listed in Table~\ref{tab:summary_GCs} should be taken with caution, especially for those clusters located at small Galactic latitudes, where \gaiahub{} could be erroneously including some MW bulge stars as members. A more detailed analysis of the possible systematic errors that could be affecting \gaiahub's results can be found in Section~\ref{sec:Systematics}.

\begin{figure}
\begin{center}
\includegraphics[width=\linewidth]{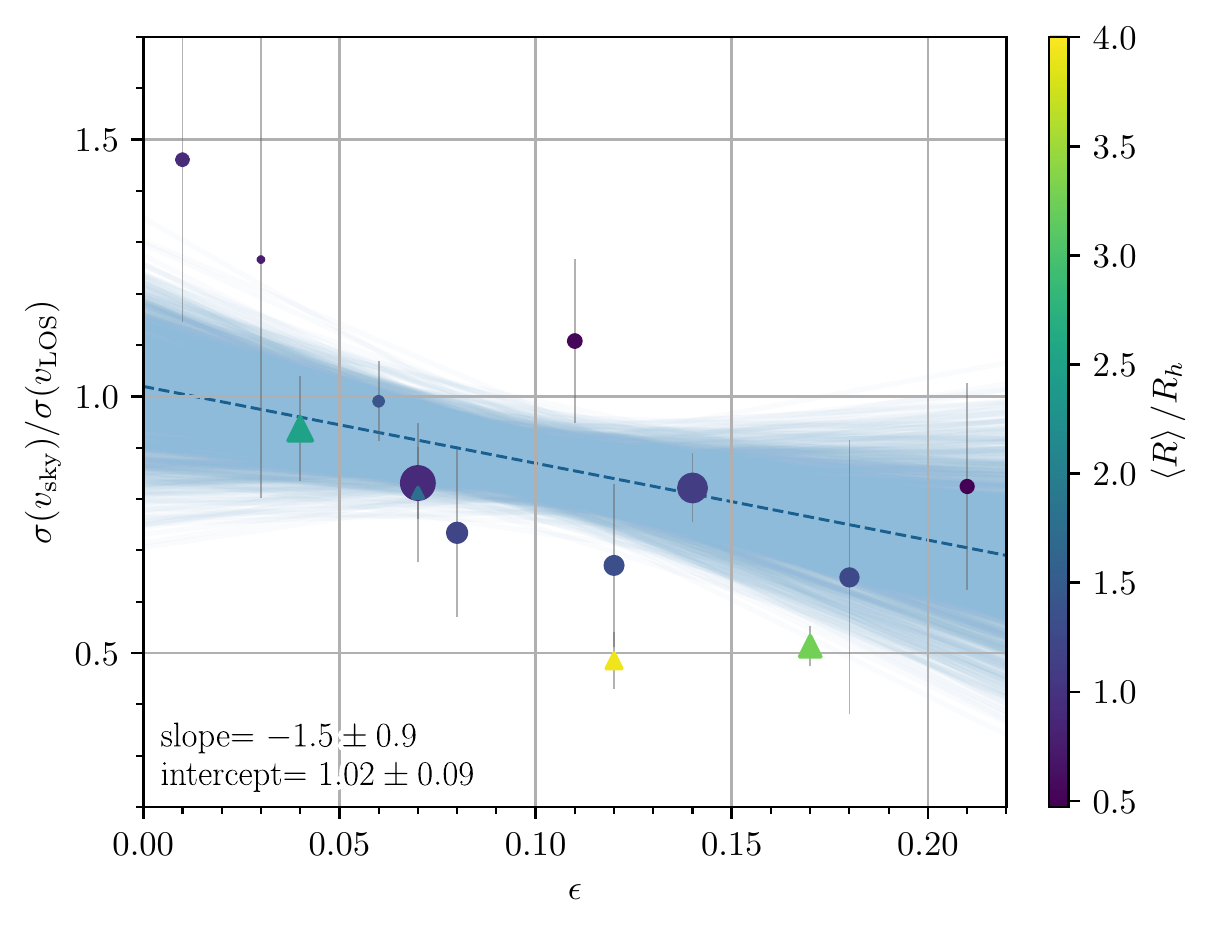}
\caption{Ratio between the sky-projected velocity dispersion, $\sigma(v_{\rm sky})$ and the line-of-sight velocity dispersion, $\svlos$, as a function of the sky-projected ellipticity $\epsilon$. Only clusters with uncertainties in $\sigma(v_{\rm sky}) / \svlos$ below 0.5 are shown. $\sigma(v_{\rm sky})$ is measured over the entire \textit{HST} field, while $\svlos$ correspond to the spectroscopic central velocity dispersion \citep[][2010 version]{Harris1996}. Clusters whose observed \textit{HST} field is located at distances larger than 1.5 times their $R_h$ are shown by triangles. The rest are shown by circles. The blue-dashed line shows the result from a straight-line fit to the clusters with $\left<R\right> / R_h < 1.5$ (circles), and highlight a mild correlation (Spearman's correlation: -0.74). The uncertainty of such fit is represented by the blue-shaded region. The rest of the markers coincide with those from Figure~\ref{fig:GCs_vrtlos}.}
\label{fig:ellipticity}
\end{center}
\end{figure}

Indeed, the ratio between the sky-projected components of the velocity dispersion, $\svr$ and $\svt$, and the line-of-sight component, $\svlos$, varies in direction and magnitude from one cluster to another. As commented above, systematic and observational effects could be behind these differences. However, the anisotropy could also be real, and be produced by the internal kinematics of the cluster. If this was the case, some correlation between the shape of the cluster and the ratio between the three components of the velocity dispersion could be present in our results. To check for this possibility, we compared the sky-projected ellipticity of the clusters ($\epsilon$) with the ratio between the average 1D dispersion observed in the plane of the sky with \gaiahub{} and the spectroscopic line-of-sight dispersion, $\sigma(v_{\rm sky})/\svlos$, where $\sigma(v_{\rm sky}) = \sqrt{ \svr^2/2 + \svt^2/2}$. Our results are shown in Figure~\ref{fig:ellipticity}, where we fit a straight line only to clusters whose \textit{HST} observations are centered on the cluster or located not far away from it ($> 1.5 R_h$, 10 clusters). As expected, we observe a mild correlation (Spearman's correlation: -0.74) between $\epsilon$ and $\sigma(v_{\rm sky})/\svlos$, with round clusters ($\epsilon \lesssim 0.1$) being, in general, more isotropic than flattened ones. Also interesting is the fact that the intercept is close to (0, 1) which indicates that perfectly round clusters should be almost perfectly isotropic. A similar behavior was found by \citet{Watkins2015} by comparing the anistropy along the projected major and minor axes. These results would suggest that the internal kinematics of these clusters is indeed shaping them and causing, at least partially, some of the differences observed between $\svr$, $\svt$, and $\svlos$.

\begin{figure}
\begin{center}
\includegraphics[width=\linewidth]{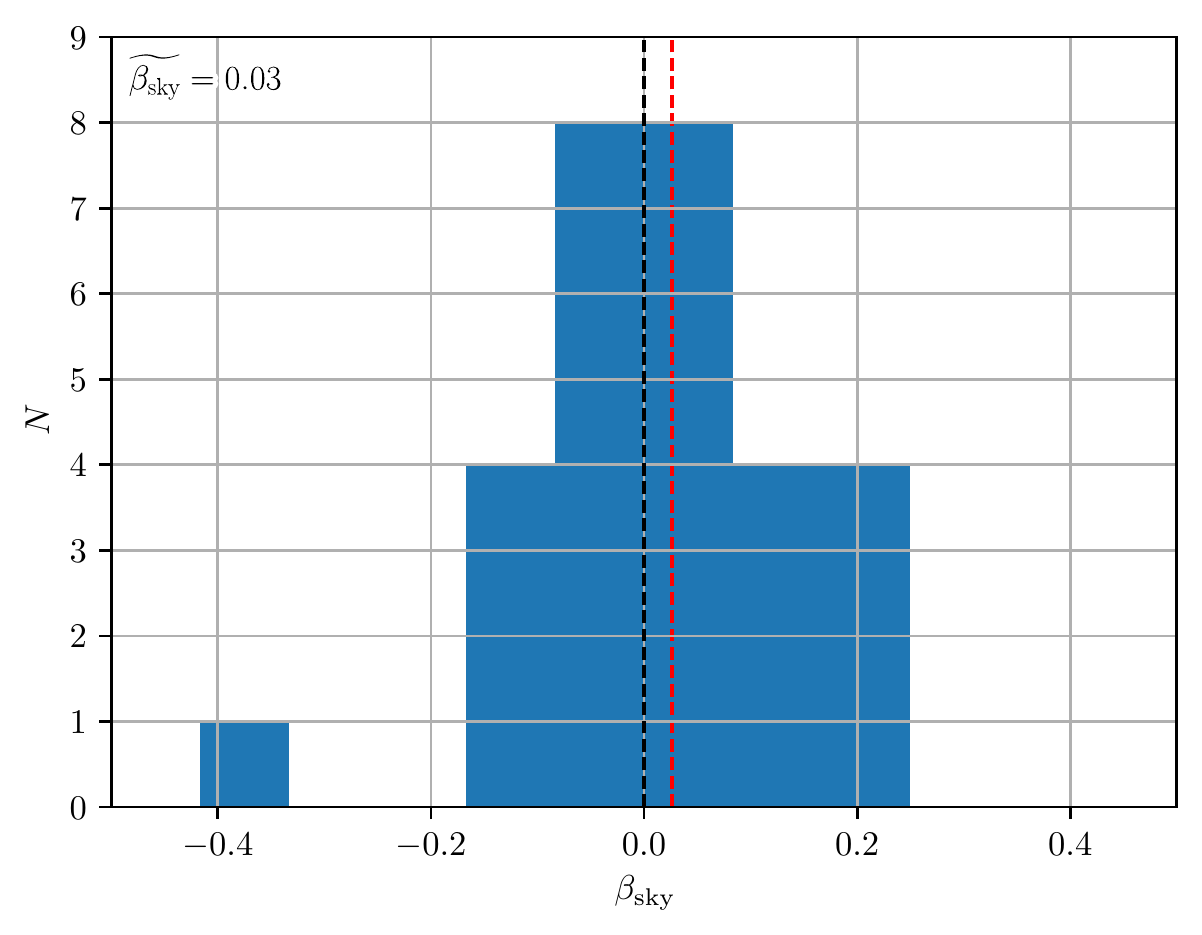}
\caption{Distribution of the anisotropy, $\beta_{\rm sky} = 1 - \svt^2 / \svr^2$, of the 37 clusters listed in Table~\ref{tab:summary_GCs}. A vertical red dashed line shows the median value of the distribution $\tilde{\beta_{\rm sky}} = 0.026$.}
\label{fig:GCs_beta}
\end{center}
\end{figure}

The distribution of anisotropy $\beta_{\rm sky} = 1 - \svt^2 / \svr^2$ for all 37 clusters is shown in Figure~\ref{fig:GCs_beta}. The distribution is slightly tilted towards radially anisotropic values, with a median value of $\tilde{\beta_{\rm sky}} = 0.026$ and an error-weighted mean value of $\overline{\beta_{\rm sky}} = 0.057 \pm 0.016$. Most of our clusters fit completely in our \textit{HST} images or have been observed at distances reaching or exceeding their half-light radii ($R_h$). Thus our results are consistent with findings of \citet{Watkins2015} that globular clusters generally become mildly radially anisotropic towards $R_h$. Similar results were also found by \citet{Vasiliev2021}, with half of the clusters in their sample showing radial anisotropy and the rest being either tangentially anisotropic or isotropic.

\subsection{Dwarf Spheroidal Galaxies}

\begin{deluxetable*}{ccccccccccc}
\tabletypesize{\scriptsize}
\tablenum{2}
\tablecaption{PM uncertainties and dispersions for dSphs using \gaiahub{} and \Gaia{} EDR3.}
\label{tab:summary_dSphs}
\tablewidth{0pt}
\tabcolsep=0.11cm
\tablehead{Name & $\dsun$ & $\Delta T$ & $\Delta(v)_\mathrm{Gaia}$ & $\Delta(v)_\mathrm{\gaiahub}$ & $\spmr$ & $\spmt$ & $\svlos$ &  $\svr$ & $\svt$ & $\beta_{\rm sky}$ \\
                  &   (kpc) & yr & $(\kms)$ & $(\kms)$ & $(\masyr)$  & $(\masyr)$ & $(\kms)$  & $(\kms)$  & $(\kms)$  &   }
\decimalcolnumbers
\startdata
Draco & $76 \pm 6$ & 12.6 & 167.0 & 11.9 & $0.025 \pm 0.006$ & $0.016 \pm 0.007$ & $9.1 \pm 1.2 (9.1 \pm 1.2)$ & $9.0 \pm 2.3$ & $5.8 \pm 2.6$ & $0.59\pm0.42$ \\
Sculptor & $86 \pm 6$ & 14.7 & 136.1 & 8.9 & $0.0202 \pm 0.0033$ & $0.018 \pm 0.004$ & $9.2 \pm 1.4 (8.8 \pm 1.8)$ & $8.2 \pm 1.4$ & $7.1 \pm 1.7$ & $0.25\pm0.43$ \\
Sextans & $86 \pm 4$ & 12.9 & 409.0 & 35.2 & $0.08 \pm 0.05$ & $0.072 \pm 0.032$ & $7.9 \pm 1.3 (6.8 \pm 2.4)$ & $32 \pm 17$ & $29 \pm 13$ & $0.16\pm1.32$ \\
Fornax & $147 \pm 12$ & 14.2 & 243.6 & 16.1 & $0.012 \pm 0.005$ & $0.018 \pm 0.006$ & $11.7 \pm 0.9 (9.2 \pm 1.8)$ & $8.0 \pm 3.5$ & $13 \pm 4$ & $-1.54\pm2.71$ \\
\enddata
\tablecomments{Columns (2) and (8) are taken from \citet{McConnachie2012}. Values in parentheses in column (8) are measured within the area covered by this study using different data available from the literature \citep{Walker2009, Walker2015}. Columns (4) and (5) are the median error in the velocity measured from PMs at $G=20$ with \Gaia{} EDR3 and \gaiahub, respectively. Column (11) shows the sky-projected anisotropy with two significant figures.}
\end{deluxetable*}

DSphs are frequently invoked as testbeds of the nature of dark matter (DM), particularly concerning the presence or absence of cusps as predicted by $\Lambda$CDM. Line-of-sight velocity surveys have suggested a generally low, nearly constant density in the cores of these galaxies
\citep[e.g.][]{Battaglia2008, Walker2011, Amorisco2012, Brownsberger2021}, which could point to either strong effects of baryonic feedback or alternative theories of DM. On the other hand, some authors found profiles that are fully consistent with $\Lambda$CDM expectations \citep{Strigari2010}, or at least compatible with both scenarios \citep{Genina2018}. Such variety of results is partly fueled by strong model degeneracies remaining in the mass profile due to the lack of accurate tangential velocity data \citep{Strigari2018, Read2021}.

\gaiahub{} can aid in this particular problem by providing tangential velocities for stars in some of the dSph satellites of the MW with accuracies below the internal velocity dispersion. Here, we run \gaiahub{} on four of the classical dSph galaxies, and provide a preliminary assessment of the quality of the results. In Table~\ref{tab:summary_dSphs}, we summarize some basic properties of the galaxies, and the derived random uncertainties for stars in the fields analyzed with \gaiahub{} at $G=20$. In three of the galaxies, Draco, Sculptor, and Fornax, \gaiahub{} manages to derive individual PMs with accuracies below the central $\svlos$ of the galaxies. The velocity dispersions are estimated following the same maximum likelihood approach that we used for the GCs. Below we analyze the results in more detail.

\subsubsection{Detailed results}

\paragraph{Draco dSph}

We derived PMs for stars in the Draco dSph, following the examples described in the Appendix~\ref{Apx:GaiaHub_Execution}. Specifically, we used the automatic membership selection in order to use only member stars to make the epoch alignment and increased the membership selection clipping probability from $3\sigma$ to $5\sigma$\footnote{This behavior is achieved by using the flags \mbox{\tt --use\_members} and \mbox{\tt --clipping\_prob\_pm} 5.}. At the time of writing this paper, \gaiahub{} found 77 suitable \textit{HST} images in Draco arranged across four fields. Three fields located close to the center of the galaxy and a more distant field with just one image and very few stars. We chose to use the images from the three central fields (GO-10229 \& GO-10812).

\begin{figure*}
\begin{center}
\includegraphics[width=\linewidth]{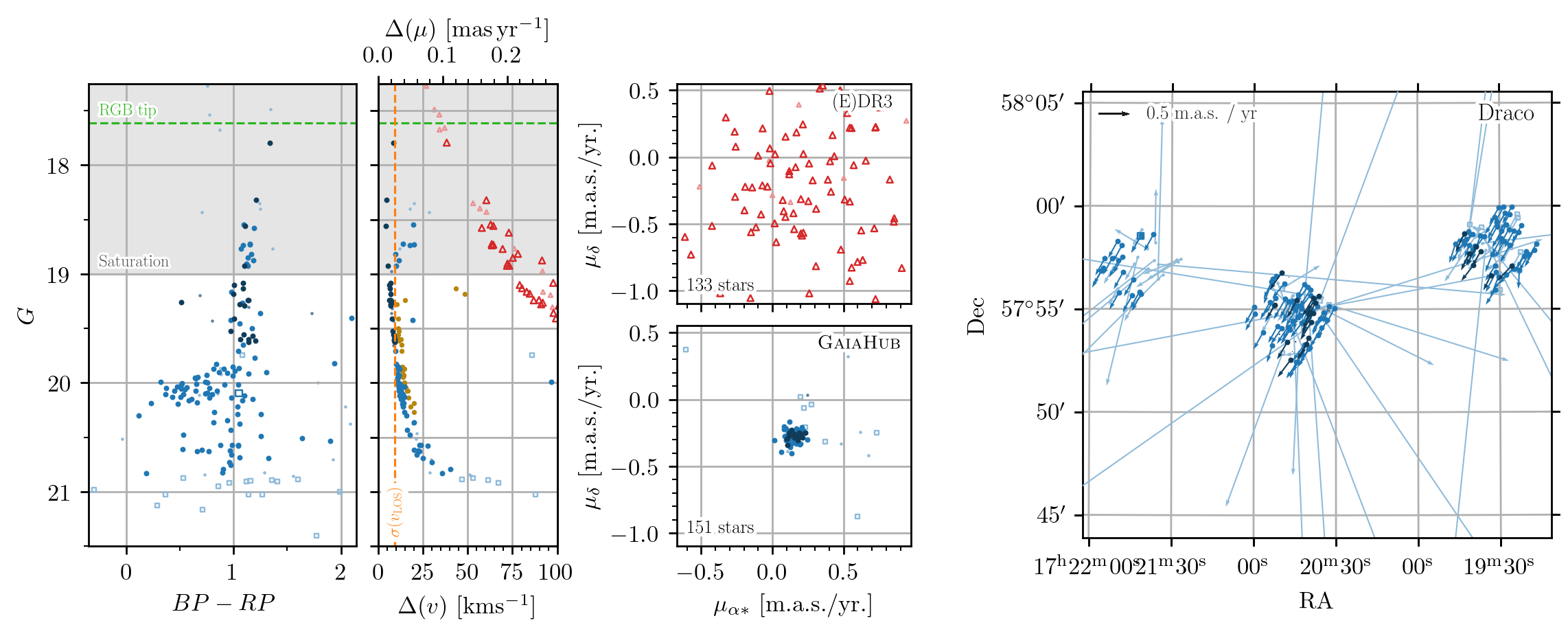}
\caption{Summary of the results for the Draco dSph galaxy. Markers and colors coincide with those of Figure~\ref{fig:NGC6535_vpd}. Results from \citet{Massari2020} are shown in the second panel for comparison by gold colored dots.}
\label{fig:Draco_vpd}
\end{center}
\end{figure*}

As described in Section~\ref{sec:reference_frame} and Appendix~\ref{Apx:Membership_selection}, after downloading all the images, \gaiahub{} runs a first iteration using all the stars available to establish a common reference frame and derive the PMs. Then, it uses the PMs to select co-moving stars, i.e. members of Draco, and repeats the process using only those to establish the reference frame. The process converged after 3 iterations, using 127 stars for the alignment between epochs and providing PMs for 151 stars. The results are summarized in Figure~\ref{fig:Draco_vpd}.

In the case of Draco, \gaiahub{} is able to derive individual PMs with uncertainties below $\svlos = 9.1 \kms$ for 21 stars. The number of member stars between the TRGB and $G=20$ is 64, with a median velocity error of $9.1\kms$. The effect of having such a small uncertainties compared with \Gaia{} alone can be observed in the concentration of the stars in the VPD, with member stars clustering much more tightly in the \gaiahub{} results. \gaiahub{} also manages to derive PMs for up to 18 stars without previous EDR3 PMs, 6 of them with uncertainties below $100\kms$.

Draco appears to be radially anisotropic, with $\beta_{\rm sky} = 0.6 \pm 0.4$, $\svr = 9.0\pm2.3 \kms$ and $\svt = 5.8 \pm 2.7 \kms$. We computed the line-of-sight velocity dispersion within the area covered in this work using radial velocity measurements from \citet{Walker2015}. The obtained value, $\svlos' = 9.1 \pm 1.2 \kms$, is based on 39 common stars, and coincides with values found in the literature for the central $\svlos$ of Draco \citep{McConnachie2012}. We used this value to estimate the intrinsic anisotropy $\beta = 1 - \svt^2/[\svlos^2 + \svr^2 - \svt^2]$ \citep[Equation 1 in][]{Massari2017} at the distance of the observed fields. Our results, $\beta = 0.75\pm0.30$, are compatible with those from \citet{Massari2020} who found $\svr = 11.0_{-1.5}^{+2.1} \kms$, $\svt = 9.9_{-3.1}^{+2.3} \kms$ and a 3D radial anisotropy of $\beta = 0.25_{-1.38}^{+0.47}$. 

\paragraph{Sculptor dSph}

As with Draco, we used the automatic membership selection and increased the membership selection clipping probability from $3\sigma$ to $5\sigma$. At the time of writing this paper, \gaiahub{} found 11 suitable \textit{HST} images in Sculptor arranged in two fields (from \textit{HST} program GO-9480). We chose to use all of the images. 

\begin{figure*}[ht]
\begin{center}
\includegraphics[width=\linewidth]{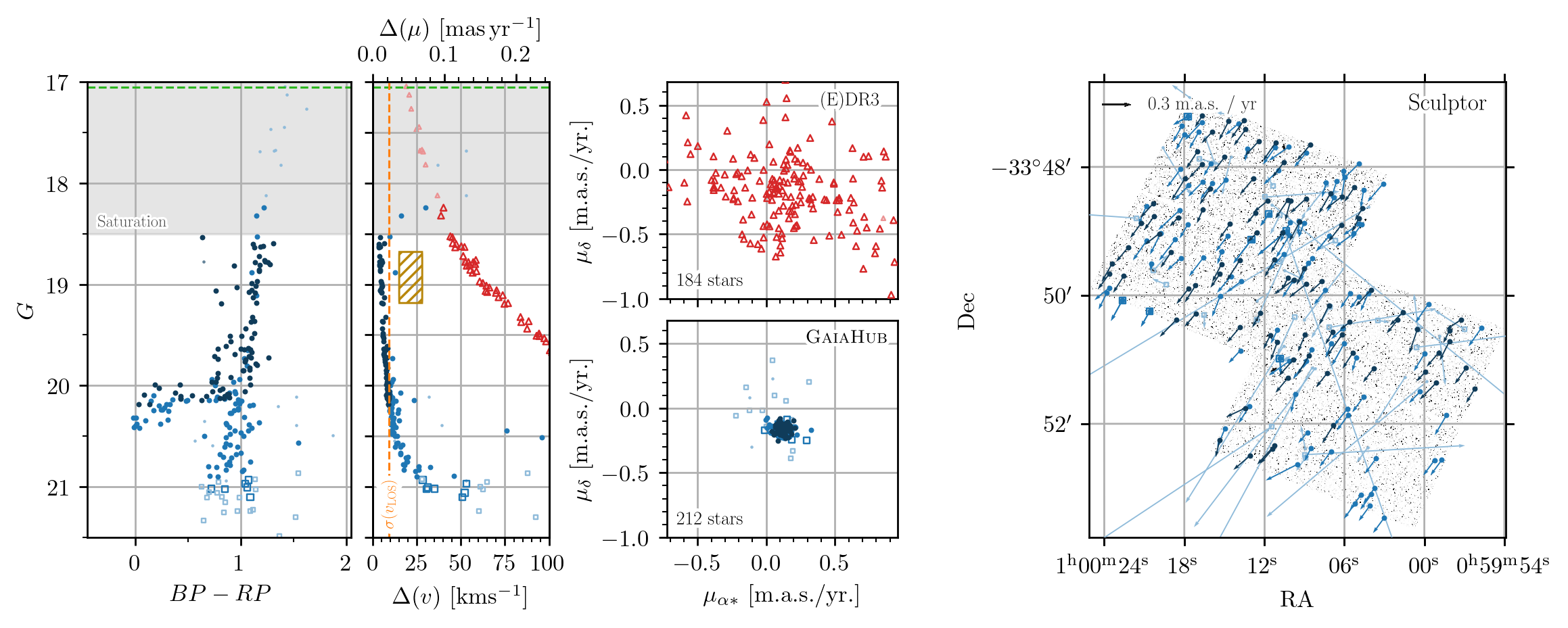}
\caption{Summary of the results for the Sculptor dSph galaxy. Markers and colors coincide with those of Figure~\ref{fig:NGC6535_vpd}. For comparison, we show the reported PM uncertainty range by \citet{Massari2018} as a gold colored hatched box in the second panel.}
\label{fig:Sculptor_vpd}
\end{center}
\end{figure*}

In the particular case of Sculptor, the process of selecting group members did not converge but started to oscillate between two possible solutions after 3 iterations. We decided to stop the execution after 4 iterations\footnote{The maximum number of iterations can be controlled using $\mbox{\tt --max\_iterations}$ or $\mbox{\tt --ask\_user\_stop}$.}, which produces the solution for which more stars labeled as members and uses 173 stars for the alignment between epochs. \gaiahub{} produced PMs for 225 stars, $\sim 14$\% more than \Gaia{} alone (197 stars). The results are summarized in Figure~\ref{fig:Sculptor_vpd}.

In the case of Sculptor, \gaiahub{} is able to derive individual PMs with uncertainties below the $\svlos$ for 81 stars. The number of member stars between the TRGB and $G=20$ is 89, with a median velocity uncertainty of $6.8\kms$. Up to 28 new PMs for stars without measurements in the EDR3 catalog were derived by \gaiahub, 13 of them with uncertainties below $100\kms$.

Using \gaiahub's PMs and assuming a distance of $\dsun =  85.9\pm5.5$ \citep{McConnachie2012}, we find a velocity dispersion in the radial and tangential direction of $\svr = 8.2\pm 1.4 \kms$ and $\svt = 7.1\pm 1.7 \kms$, respectively (propagating the error in distance). These values, are in good agreement with those derived by \cite{Massari2018}, where combining \textit{HST} and \Gaia{} DR1 data they found $\svr = 11.5\pm 4.3 \kms$ and $\svt = 8.5\pm 3.2 \kms$. This would indicate that Sculptor is radially anisotropic at the position of the observed \textit{HST} fields, amid a large relative uncertainty in the measurement ($\beta = 0.25 \pm 0.43$). As with Draco, we use $\vlos$ measurements from \citet{Walker2009} to estimate $\svlos' = 8.8\pm1.8 \kms$ in the observed region (based on 15 stars). The value for the intrinsic anisotropy, $\beta = 0.46\pm0.44$, also indicates that Sculptor is mildly radially anisotropic.

\paragraph{Sextans dSph}

\begin{figure*}
\begin{center}
\includegraphics[width=\linewidth]{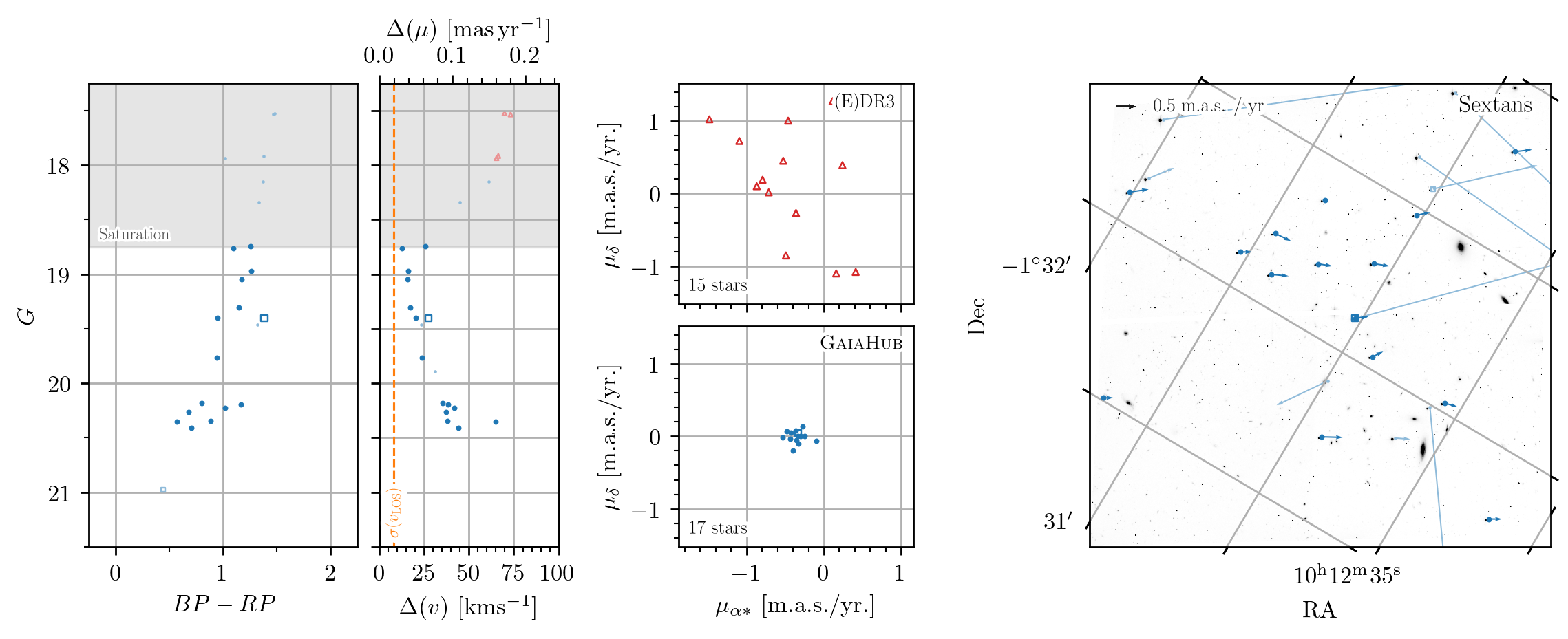}
\caption{Summary of the results for the Sextans dSph galaxy. Markers and colors coincide with those of Figure~\ref{fig:NGC6535_vpd}.}
\label{fig:Sextans_vpd}
\end{center}
\end{figure*}

Two suitable \textit{HST} fields were found for Sextans. However, due to the scarce number of stars, and the low ratio between members and foreground MW's stars, results for only one field converged and produced PMs (GO-10229, 15 images). Results are summarized in Figure~\ref{fig:Sextans_vpd}. In total, 17 stars were measured in the field, with 15 of them being classified as members. For these stars, \gaiahub{} yielded far more precise results than those of \Gaia{} alone, with uncertainties around $35 \kms$ at $G=20$ compared to uncertainties of $\sim 400 \kms$ for \Gaia. However, the scarce number of measured PMs and their relatively large uncertainties compared to the central velocity dispersion of Sextans ($\svlos = 7.9 \pm 1.3$), did not allow us to obtain statistically sound results for the dispersion velocity. The derived values, $\svr = 32 \pm 17$ and $\svt =29 \pm 13$, are far too high compared with $\svlos$. This could also indicate that non-negligible systematic effects are affecting our data, but with only 15 member stars, our tests remained inconclusive (see Section~\ref{sec:Systematics}).

\paragraph{Fornax dSph}

\begin{figure*}
\begin{center}
\includegraphics[width=\linewidth]{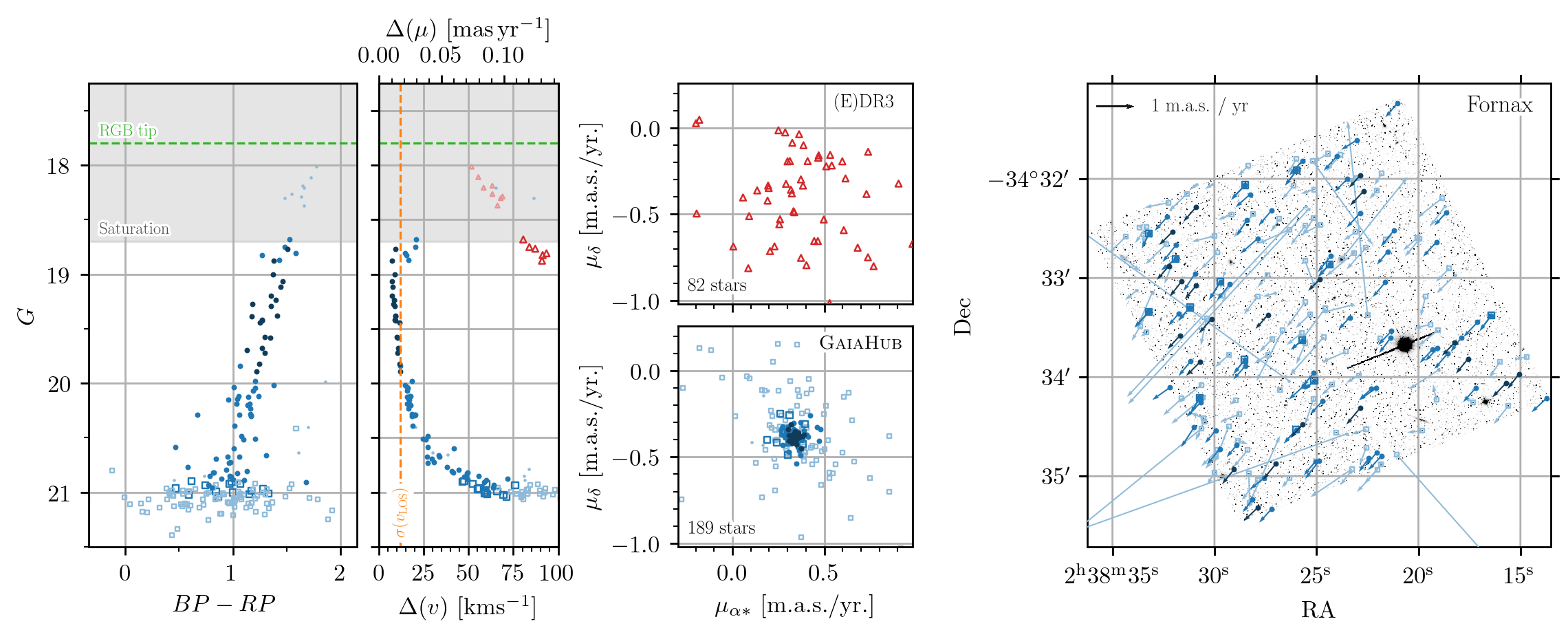}
\caption{Summary of the results for the Fornax dSph galaxy. Markers and colors coincide with those of Figure~\ref{fig:NGC6535_vpd}.}
\label{fig:Fornax_vpd}
\end{center}
\end{figure*}

\gaiahub{} found 6 suitable \textit{HST} images in Fornax arranged in a single field (GO-9480 and GO-9575). We chose to use all of the images. \gaiahub{} converged after two iterations, using 54 stars for the alignment between epochs and providing PMs for 198 stars. The results are summarized in Figure~\ref{fig:Fornax_vpd}.

Despite its large heliocentric distance ($\dsun = 147 \pm 12 \kpc$), \gaiahub{} is able to derive individual PMs with uncertainties below the $\svlos = 11.7\pm0.9 \kms$ for 21 stars in Fornax. The number of member stars between the TRGB and $G=20$ is 38, with a median velocity uncertainty of $11.6\kms$ and a typical uncertainty at $G=20$ of $16.1\kms$. \gaiahub{} also manages to derive PMs for up to 107 stars without previous EDR3 PMs, 39 of them with uncertainties below $100\kms$. As with previous examples, the gain in precision with respect to \Gaia{} EDR3 results is easily noticeable in the VPD, with member stars clustering much more tightly on the \gaiahub's VPD.

The obtained velocity dispersion, $\svr = 8.0\pm 3.5 \kms$ and $\svt = 13\pm 4 \kms$ would suggest that Fornax is tangentially anisotropic at the location of the HST fields. However, the relatively large uncertainties of our measurements do not allow us to reach any strong conclusion ($\beta = -1.5 \pm 2.7$). Measurements of the dispersion along the line-of-sight ($\svlos' = 9.2\pm1.8$, based on 16 stars) do not allow for any improvement in Fornax, with uncertainties that make our results also, compatible with both radially and tangentially anisotropic.

\subsubsection{Anisotropy and caveats}

\gaiahub{} provided far more precise results than those of \Gaia{} alone in all the cases analyzed in this work. Our results show that the general tendency among the dSphs satellites of the MW is to be radially anisotropic, with the exception of Fornax. This is compatible with findings of \citet{Massari2018, Massari2020}, who used the same instruments and similar methods but previous \Gaia{} data releases, and hence shorter time baselines.

Despite the clear improvement, our results are still affected by relatively large uncertainties that prevent us from making any strong claim about the anisotropy of the dSphs, except for Draco and perhaps Sculptor. Nevertheless, it is interesting that Fornax is the only system that appears to be tangentially anisotropic (amid very large uncertainties). Fornax exhibits relatively complex rotation patterns both along the line-of-sight \citep{delPino2017} and on the plane of the sky \citep{Martinez-Garcia2021}. It is also the only galaxy in our sample that host GCs and known stellar shells. This, among other conspicuous features of its star formation history, has led some authors to claim that Fornax could be the remnant of a merger \citep{Amorisco2012, delPino2015}. If this was the case, the galaxy's peculiar kinematics could be behind its possible tangential anisotropy. Future \Gaia{} releases will increase both the positional accuracy and the time baseline, which will greatly improve the uncertainties and shed light on these questions.

\section{Systematic errors}\label{sec:Systematics}

Both observatories, \Gaia{} and \textit{HST}, show undesired systematic errors that affect their astrometric measurements. In the case of \textit{Gaia}, the magnitude and direction of these systematics are known to be dependent on the considered position in the sky, the apparent brightness of the stars and their color. While some characterization of these systematics exists, \citep[for example][]{Lindegren2021, Fardal2021, Vasiliev2021}, correcting for their effects has turned out to be difficult. Moreover, the behavior of systematics on angular scales as small as an HST field are effectively unconstrained. A solution some authors adopt is to offer alternative astrometric zero-points measured using stationary sources (distant quasars) located within a few degrees around the object of interest \citep{Marel2019, delPino2021, Martinez-Garcia2021, Battaglia2021}. For \textit{HST}, the positional accuracies for stars are affected by several factors, most importantly the geometric distortions affecting the focal plane, CCD charge transfer inefficiency, and {\it breathing} of the telescope (i.e. the expansion or contraction of the observatory due to heating by solar radiation). These effects are all corrected by the \textit{HST} data-processing pipeline and/or the data reduction performed by \gaiahub. However, the corrections are never perfect, and residual systematics may remain. The combination of the two observatories can therefore be affected by systematic errors that are difficult to characterize.

One way to estimate the systematics is to analyze the velocity dispersion $\sigma(\mu)$ derived from PMs, and compare it to other independent measurements. In general, $\sigma(\mu)$ is the convolution of the intrinsic PM dispersion of the system, $\sigma_{int}(\mu)$, the random uncertainties, $\Delta(\mu)$, and the systematics, $\Delta_{sys}(\mu)$. Therefore, systematic uncertainties could be derived as $\Delta_{sys} = \sqrt{\sigma(\mu)^2 - \Delta(\mu)^2 - \sigma_{int}(\mu)^2}$. While independent measurement of $\sigma_{int}(\mu)$ mostly do not exist at the moment, measurements of the central velocity dispersion along the line-of-sight, $\svlos$ are fairly common. By assuming that the internal dispersion of the stellar object is approximately isotropic in all directions, we can substitute $\sigma_{int}(\mu)$ by $\svlos$ in the relation above to infer systematic errors. The best objects to try this method are GCs, whose sphericity suggests that both quantities should indeed be similar. 

\begin{figure*}
\begin{center}
\includegraphics[width=\linewidth]{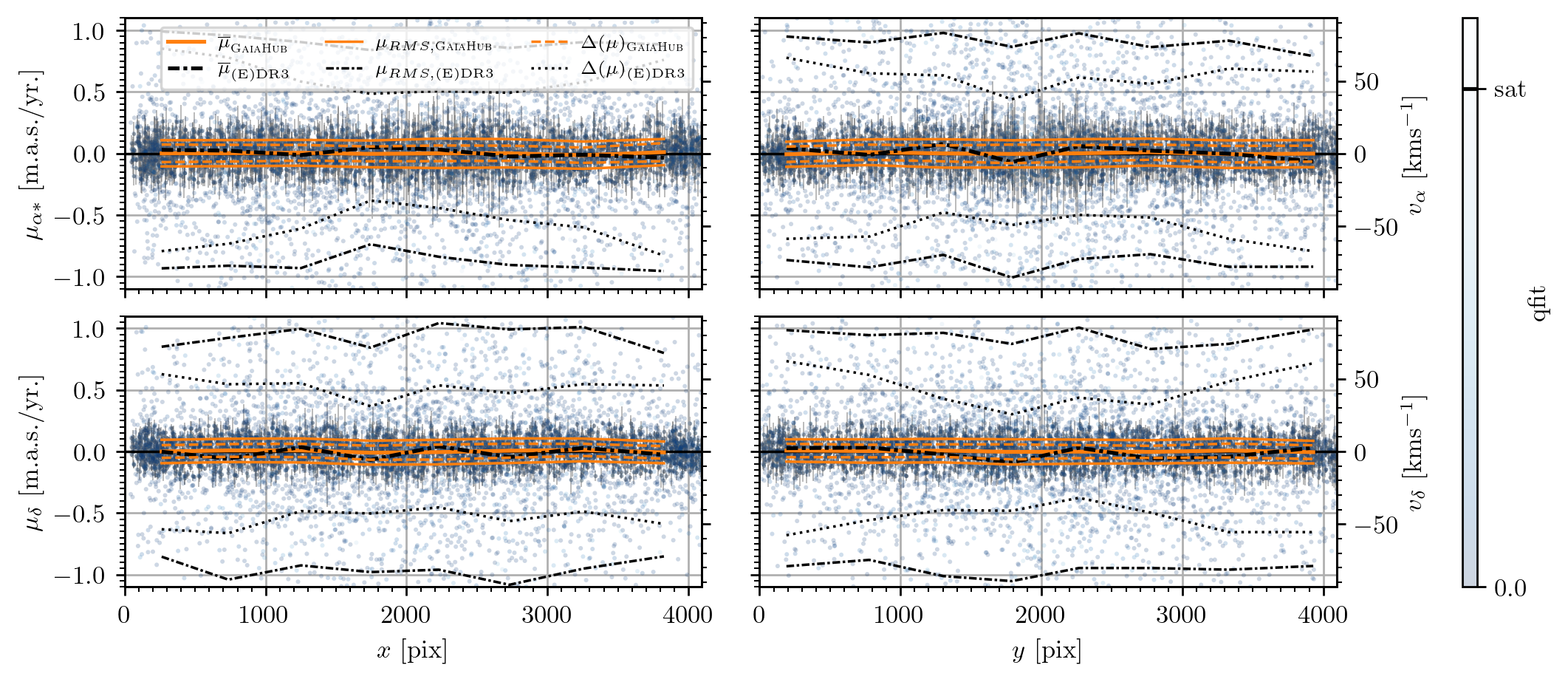}
\caption{Relative stellar PMs (and velocities) in NGC\ 5024 as a function of the CCD's position in the \textit{HST} images. The mean value of $\pmra$ and $\pmdec$ ($\overline{\mu}$), the RMS ($\mu_{\rm RMS}$), and the mean random error ($\Delta(\mu)$) are shown by thick, thin, and thin-dashed orange lines, respectively, in the case of \gaiahub{}. For comparison, the same statistics are shown for \Gaia{} by thick, thin, dotted-dashed and dotted lines in black color. All statistics are computed within 8 quantiles along both CCD coordinates $(x, y)$. The color scale indicates the mean value of goodness of the PSF fit of the star in the HST data (qfit). Stars with colors lighter than the qfit = "sat" are saturated in the \textit{HST} images. Only member stars show error bars. The large difference between the RMS and the random uncertainties cannot be explained by the internal velocity dispersion of the cluster and may indicate the presence of systematic errors.}
\label{fig:sys_xy_NGC5024}
\end{center}
\end{figure*}

As a demonstration we analyze NGC\ 5024, a cluster that we suspect suffers from systematic errors due to the high stellar crowding observed in its central region. This may be behind the relatively large values of $\svr = 6.6 \pm 1 \kms$ and $\svt = 6.3\kms$, $\sim 2 \sigma$ above the central velocity dispersion along the line-of-sight, $\svlos = 4.4\pm0.9 \kms$. Figure~\ref{fig:sys_xy_NGC5024} shows the relative PMs along the $x$ and $y$ axes of the \textit{HST} CCD for NGC\ 5024. Results from \gaiahub{} are remarkably stable compared to those of \Gaia{} alone, with the average relative PMs, $\overline{\mu}$ (orange line) forming a straight line centered at zero along both directions in the CCD. However, for this particular cluster, $\mu_{\rm RMS}$ (orange solid thin lines) is consistently larger than $\Delta(\mu)$ (orange dashed thin lines). This difference cannot be explained by its internal dispersion alone if we assume that NGC\ 5024 is isotropic. Applying the expression from the previous paragraph, we obtained values of $\Delta_{sys}(\pmra) \sim 71 \uasyr$ and $\Delta_{sys}(\pmdec) \sim 56 \uasyr$, equivalent to $\sim 6$, and $\sim 5 \kms$, respectively ($\sim 0.79$ and $\sim 0.62 \mas$ if we multiply by $\Delta(T) = 11.24$ years for NGC 5024). Indeed, if we add in quadrature these values to the PMs nominal uncertainties and repeat our calculations for the sky-projected dispersions, we obtain $\svr = 4.7 \pm 0.6 \kms$ and $\svt = 4.5 \pm 0.6 \kms$, values that are fully compatible with $\svlos$. It is important to point out that, because this method assumes that the cluster is isotropic, i.e. $\svr \sim \svt \sim \svlos$, the fact that adding these values to the final error budget results in compatible dispersion values along the three dimensions should not be interpreted as these being an accurate measurement of the systematics.

Repeating this calculation for all the GCs with $\sigma(\pmra)$ or $\sigma(\pmdec)$ greater than $\svlos$ from our sample yields a median, all-sky $\widetilde{\Delta_{sys}}(\pmra, \pmdec) \sim (25, 15)\uasyr$ ($\sim 0.27$ and $\sim 0.16 \mas$ when multiplying by the temporal baseline). As commented above, in reality the clusters might well be non-isotropic (see Section~\ref{sec:GCs_anisotropy}), and therefore these values should serve only as an estimation of the typical maximum systematic errors currently affecting \gaiahub's results.

Another way to assess the impact of systematics is to analyze the RMS scatter observed in the PMs ($\mu_{\rm RMS}$) and compare it to their random uncertainties, $\Delta(\mu)$. In clusters where the intrinsic velocity dispersion is low, $\mu_{\rm RMS}$ and $\Delta(\mu)$ should be similar unless the effects from systematics are not negligible. The cluster NGC\ 5053 seems to be an ideal example target; it shows a very low velocity dispersion, $\svlos = 1.4\pm0.2$, has a large number of stars observed by \textit{Gaia}, and is located at a large distance ($\dsun = 17.4 \kpc$), where systematics should dominate the total uncertainty budget. Because of the relatively large distance to the cluster and its low line-of-sight velocity dispersion, only one star has PMs uncertainties below the $\svlos$ value. However, the results are clearly better than those of \Gaia{} alone (see Figure~\ref{fig:NGC5053_vpd}), and allow us to perform a detailed analysis of the distribution of the PMs and their random uncertainties as a function of different parameters.

\begin{figure*}
\begin{center}
\includegraphics[width=\linewidth]{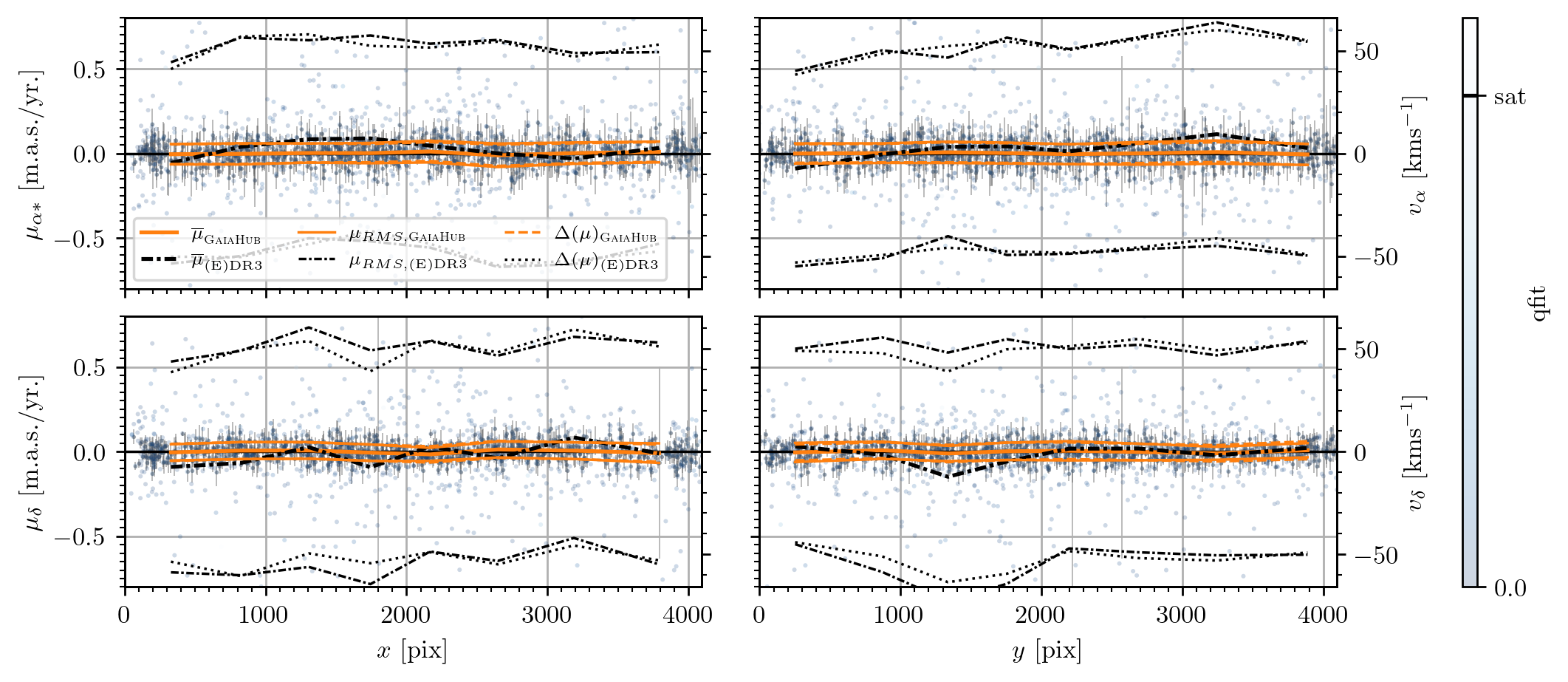}
\caption{Relative stellar PMs in NGC\ 5053 as a function of the CCD's position in the \textit{HST} images. Markers and colors coincide with those of  Figure~\ref{fig:sys_xy_NGC5024}. Notice that in this case, the RMS and the random uncertainties are almost identical, indicating that systematics do not have a large impact on the results.}
\label{fig:sys_xy_NGC5053}
\end{center}
\end{figure*}

Figure~\ref{fig:sys_xy_NGC5053} shows the relative PMs along the $x$ and $y$ axes of the \textit{HST} CCD for NGC\ 5053. As with NGC\ 5024, results obtained for NGC\ 5053 with \gaiahub{} are far more stable than those from \Gaia{} alone. However, in this case $\mu_{\rm RMS}$ is very similar to $\Delta(\mu)$, which would indicate that systematics errors have a negligible effect on the derived PMs.

\begin{figure*}
\begin{center}
\includegraphics[width=\linewidth]{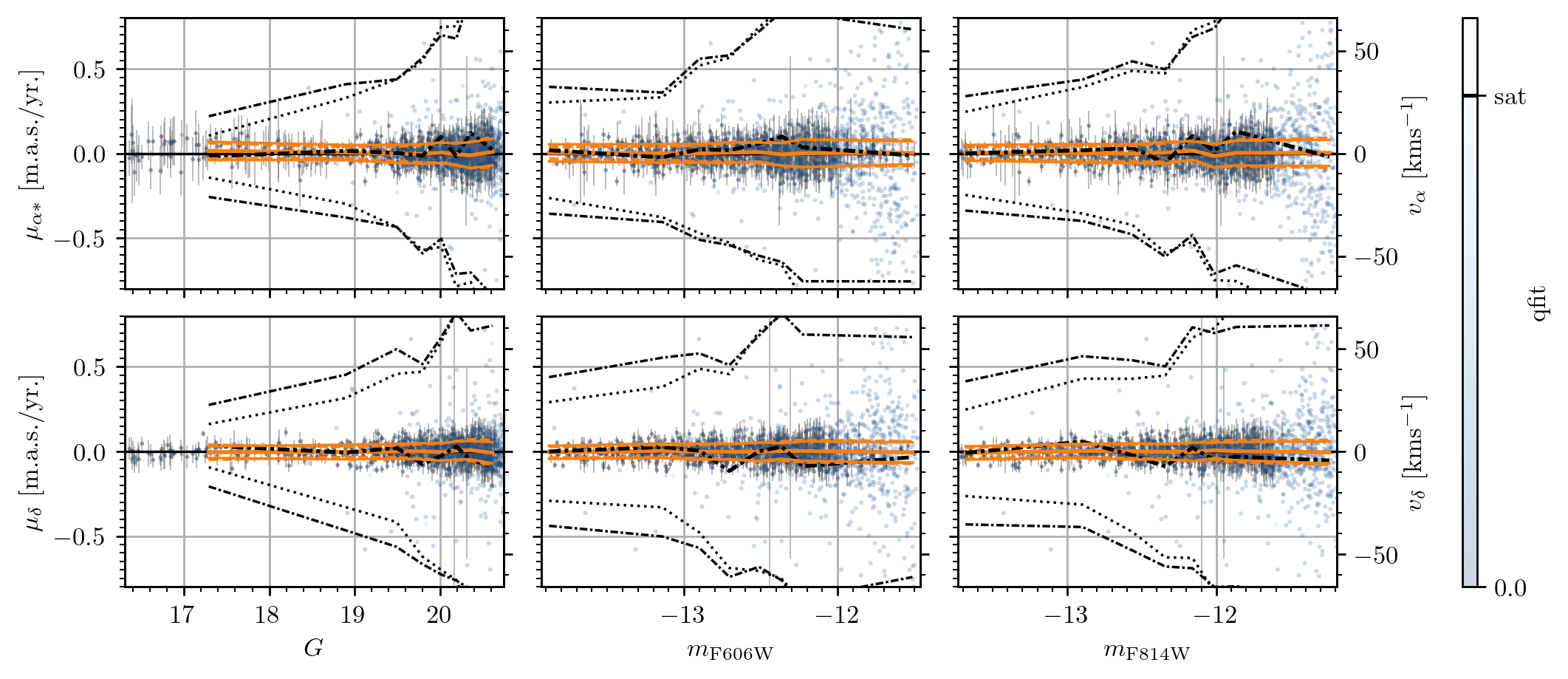}
\caption{The relative PMs of NGC\ 5053 as a function of magnitude. Markers and colors coincide with those of  Figure~\ref{fig:sys_xy_NGC5024}}
\label{fig:sys_mag_NGC5053}
\end{center}
\end{figure*}

\begin{figure*}
\begin{center}
\includegraphics[width=\linewidth]{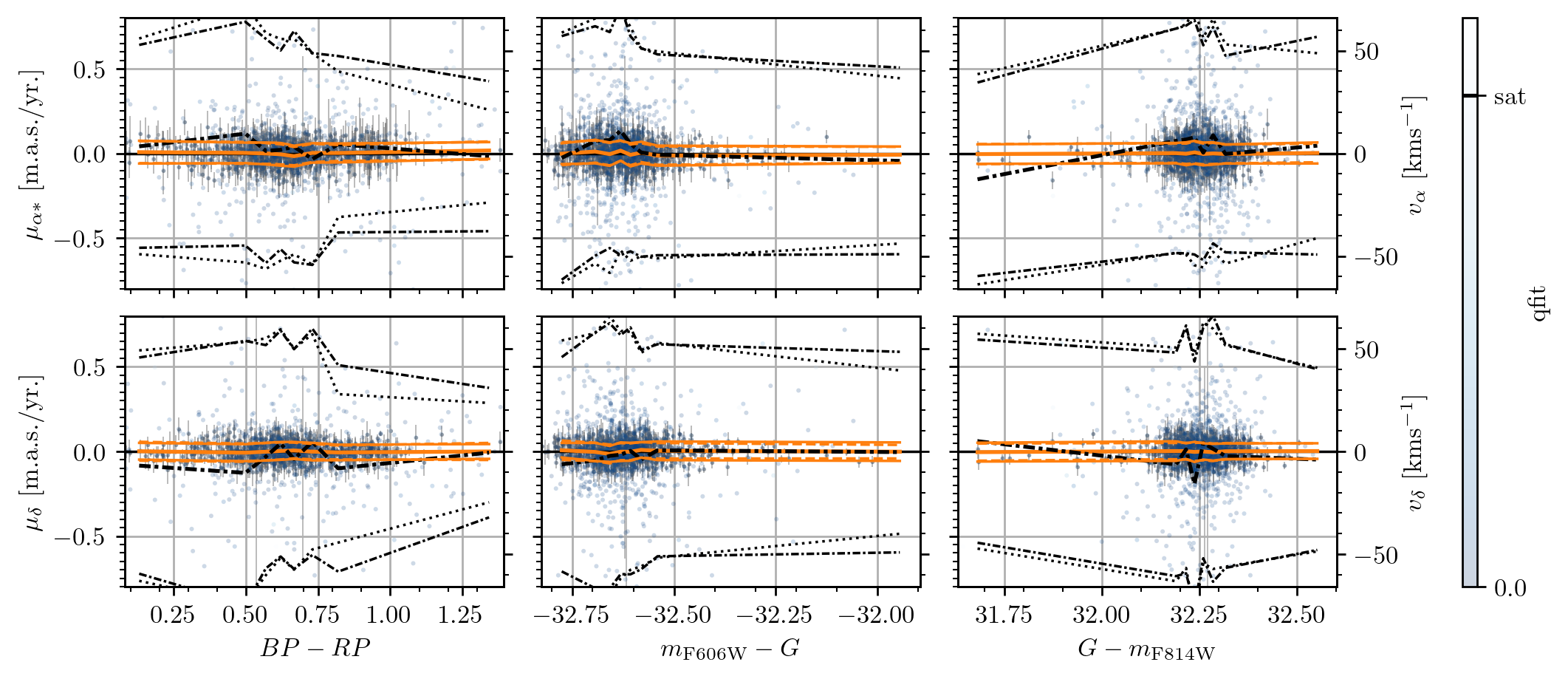}
\caption{The relative PMs of NGC\ 5053 as a function of color. Markers and colors coincide with those of  Figure~\ref{fig:sys_xy_NGC5024}.}
\label{fig:sys_col_NGC5053}
\end{center}
\end{figure*}

The same relative PMs are shown as a function of the stellar magnitudes, and color in Figures~\ref{fig:sys_mag_NGC5053} and~\ref{fig:sys_col_NGC5053}, respectively. In general, \gaiahub's results outperforms those from \Gaia{} in stability and precision, and show mean random uncertainties that are very close to the observed RMS. Assuming that $\sigma_{int}(\mu) \approx \svlos$, we can calculate the systematics as $\Delta_{sys}(\mu) = \sqrt{\sigma(\mu)^2 - \Delta(\mu)^2 - \svlos^2}$. In the case of NGC\ 5053, this yields systematic errors on the order of $\Delta_{sys}(\pmra) \sim 13 \uasyr$ and $\Delta_{sys}(\pmdec) \sim 10 \uasyr$ ($\sim 0.15$ and $\sim 0.11$ mas with $\Delta(T) = 11.3$ years for NGC 5053).

Systematic errors are expected to have contributions from \HST{}, \Gaia{}, and from the epoch alignment procedure itself, which makes disentangling their origin far more complicated than trying to measure their impact on the final PMs. However, we noticed that the systems that seem to be more affected by systematics show larger differences between $\mu_{\rm RMS, {\rm (E)DR3}}$ and $\Delta(\mu)_{\rm (E)DR3}$, regardless of their internal dispersion. This can be seen, for example, by comparing both curves in Figures~\ref{fig:sys_xy_NGC5024} and ~\ref{fig:sys_xy_NGC5053}. In the case of NGC\ 5053, both curves are practically one on top of the other, while in NGC\ 5024 $\Delta(\mu)_{\rm (E)DR3}$ is $\sim 1.8$ times smaller than $\mu_{\rm RMS, {\rm (E)DR3}}$ and cannot account for the observed dispersion. This could indicate that the nominal errors in \Gaia{} are largely underestimated in NGC\ 5024 and therefore that \Gaia{} PMs are affected by larger systematics in this case. Moreover, the systematic error values found for NGC\ 5053 are similar to those reported for \Gaia{} EDR3 PMs at small scales ($0-0.1\degr$) for the entire sky \citep{Lindegren2021, Martinez-Garcia2021, Vasiliev2021}, which may indicate that, if present, most of the systematics errors found in \gaiahub's results are being propagated from those affecting \Gaia's stellar positions. Therefore, \gaiahub{} would not be introducing any noticeable systematic errors in the final results, and thus we expect its results to greatly improve as \Gaia's systematics drop in future data releases.

However, we should point out that this might not be the case for other stellar systems. Systematics affecting both instruments vary depending on the quality of the used \textit{HST} and \Gaia{} data, which could result in very different scenarios depending on the considered object. Furthermore, running \gaiahub{} with different options also has on impact on the results, and thus could mitigate or increase the impact of systematic errors. Finally, it is also worth noticing that the estimations provided here may be not valid with future \Gaia{} releases or increased time baselines. How to best characterize or try to correct for systematics will ultimately depends on the scientific goals of the project. Therefore, we recommend the user to exercise extreme caution and to thoroughly analyze the results before reaching any scientific conclusions.

\section{\gaiahub{} usability}\label{sec:Usability}

\subsection{When is it a good idea to use \gaiahub?}

In normal conditions, \gaiahub{} will always provide more precise PMs for faint stars than \Gaia{} alone. However, the results will be limited to the field of view of \textit{HST}, which will significantly reduce the number of observed stars in stellar systems larger than the covered area. This is the case in dSph galaxies and in some nearby and/or very massive GCs. Another aspect to consider is the limited magnitude range in which both instruments have common measurements, $17 \leq G \leq 21$ mag (brighter stars are often saturated in \textit{HST} images). This also limits the number of stars for which \gaiahub{} can derive PMs.

Taking these considerations into account, the usefulness of \gaiahub{} will generally depend on the particular scientific application. \gaiahub{} will provide better PM measurements on a star-by star basis, making it a very interesting tool to derive sky-projected velocity dispersions, or to find runaway stars. An ideal example case for \gaiahub{} would be a GC at a distance $\gtrsim 50 \kpc$; small enough so as to fit in the \textit{HST} field of view, and distant enough so its brightest RGB stars are not saturated for typical \textit{HST} exposure lengths. For dSphs, \gaiahub{} will also provide more precise PMs, but given the small coverage of \textit{HST} in these systems, their scientific usability is more limited. 

In short, \gaiahub{} is most useful for stellar systems at distances $\dsun \gtrsim 50 \kpc$. For much closer objects, \Gaia{} can use very bright stars that are saturated in \textit{HST} images, and there is little advantage to adding \textit{HST} data. For very distant objects (several hundreds of kpc), \Gaia{} detects very few stars and \textit{HST} observations alone are preferable.

\subsection{Other stellar fields and uses}

\gaiahub{} can be used in any kind of stellar field, not only stellar clusters or galaxies. However, a minimum number of at least $\sim 100$ stars is desirable to establish the reference frame. It is possible to use \gaiahub{} in less populated fields, although this might impact the quality of the results. If there is not a co-moving stellar population in the field of study, we strongly recommend running the code without automatic membership selection\footnote{Without the $\mbox{\tt --use\_members}$ flag}.

Lastly, \gaiahub{} can also be used to determine precise systemic PMs. The larger number of stars with PMs measured with \gaiahub{}, combined with the higher precision of the measurements compared to those of \Gaia{}, allows for a more precise determination of systemic PMs in distant stellar systems (Bennet et al. ApJ submitted).

\section{Conclusions}\label{sec:Conclusions}

We have presented \gaiahub, a tool that combines \textit{HST} archival images with \Gaia{} measurements to derive precise PMs. \gaiahub{} boosts the scientific impact of both observatories beyond their individual capabilities by providing a second epoch observation for any \textit{HST} archival image, and improving the PM accuracy for any faint source $(G \gtrsim 18)$ in the \Gaia{} catalog observed by \textit{HST} more than $\sim 6$ years ago. Our results show that random uncertainties with \gaiahub{} improve by roughly a factor $\Delta T_{\rm Gaia + HST}/\Delta T_{\rm Gaia}$ over those of \textit{Gaia}, which is equivalent to PMs $\sim 10$ times more precise than those of \Gaia{} EDR3 at $G = 19.5$ when using \textit{HST} observations taken in the year 2007. While the differences in precision between \Gaia{} and \gaiahub{} will drop with future \Gaia{} data releases, \gaiahub{} will always produce more precise PMs at fainter magnitudes, making it interesting for a large number of stellar systems in the Local Group. \gaiahub{} is completely public and accessible for everyone, and we plan to maintain it and update it.

As demonstration of its capabilities, we used \gaiahub{} to derive the internal PMs of 4 dSph galaxies; Draco, Sculptor, Sextans, and Fornax; and 37 globular clusters with just one \textit{HST} epoch, or located at distances larger than 25 kpc. Some of the systems are located at large distances of order $100 \kpc$. The precision achieved with \gaiahub{} allowed us to measure tangential velocities of individual stars with accuracies below the central velocity dispersion values in almost all the analyzed systems, e.g. $\Delta(\mu) \sim 1.3 \masyr$ ($\sim 9.1 \kms$) at $G = 19.5$ in Fornax ($\svlos = 11.7\pm0.9 \kms$, $\dsun = 147 \pm 12 \kpc$). We used these measurements to derive the 2D sky-projected velocity dispersion values. Our results are generally consistent with those available in the literature derived from line-of-sight velocity measurements. They are also compatible with those derived using \textit{HST}-only PMs, where available.

We confirm existing results for other samples that GCs tend towards mild radial velocity dispersion anisotropy. We also find that the shape of the GCs is related to their internal kinematics, with more round clusters being more isotropic than those showing smaller $\sigma(v_{\rm sky})/\svlos$ ratios. Lastly, we also confirm previous findings that Draco and Sculptor appear to be radially anisotropic systems. For systems such Fornax of Sextans a longer time baseline is required in order to derive more consistent results.

Lastly, we have measured the impact of possible systematic effects in \gaiahub{} results following two different approaches. Results from these tests yield an all-sky median maximum error of $\widetilde{\Delta_{sys}}(\pmra, \pmdec) \sim (25, 15)\uasyr$. We expect these systematics to improve with future \Gaia{} data releases.

{\it Acknowledgements:}
The authors thank the anonymous referee for the comments that have helped to improve this paper. Support for this work was provided by a grant for HST archival program 15633 provided by the Space Telescope Science Institute, which is operated by AURA, Inc., under NASA contract NAS 5-26555. A. del Pino acknowledges the financial support from the European Union - NextGenerationEU and the Spanish Ministry of Science and Innovation through the Recovery and Resilience Facility project J-CAVA. A. del Pino also thanks Dr. Bertran de Lis and Mr. Pi\~nero Diaz for their support and help during the realization of this project. This work has made use of data from the European Space Agency (ESA) mission \Gaia{} (\url{https://www.cosmos.esa.int/gaia}), processed by the \Gaia{} Data Processing and Analysis Consortium (DPAC, \url{https://www.cosmos.esa.int/web/gaia/dpac/consortium}). Funding for the DPAC has been provided by national institutions, in particular the institutions participating in the \Gaia{} Multilateral Agreement. This work is part of the HSTPROMO (High-resolution Space Telescope PROper MOtion) Collaboration\footnote{\url{http://www.stsci.edu/~marel/hstpromo.html}}, a set of projects aimed at improving our dynamical understanding of stars, clusters and galaxies in the nearby Universe through measurement and interpretation of proper motions from \textit{HST}, \Gaia, and other space observatories. We thank the collaboration members for the sharing of their ideas and software.

{\it Software:} 
{\code{numpy} \citep{numpy}, 
\code{scipy} \citep{scipy},
\code{matplotlib} \citep{matplotlib}, 
\code{astropy} \citep{astropy1,astropy2},
}

\appendix

\section{\gaiahub{} execution}\label{Apx:GaiaHub_Execution}

\gaiahub{} is designed as an automatic pipeline that can run at a wide variety of user input levels. It can compute and manage all the technical nuisance parameters and steps required in order to measure PMs combining \HST{} and \Gaia{} data. This includes but it is not limited to: finding and downloading suitable \HST{} images, performing astrometric measurements in these, dealing with different data quality and time baselines, dealing with saturated stars, and membership selection.

In this Appendix we provide a general guideline on how to execute \gaiahub{} and its available options. The best way to learn about these options is through the help included with \gaiahub;

\begin{verbatim}
$ gaiahub --help
\end{verbatim}.

\subsection{Basic execution}\label{Apx:Basic_execution}

\gaiahub{} allows users to automatically search the MAST\footnote{\url{https://mast.stsci.edu/portal/Mashup/Clients/Mast/Portal.html}} and Gaia\footnote{\url{https://gea.esac.esa.int/archive/}} catalogs for suitable \textit{HST} images and \Gaia{} stars around a set of coordinates in the sky;

\begin{verbatim}
$ gaiahub --ra 15.03898 --dec -33.70903
\end{verbatim}.

Since no search radius was provided in the example above, \gaiahub{} will ask the user what radius they want to use. Another option is to search by the name of the object of interest;

\begin{verbatim}
$ gaiahub --name "Sculptor dSph"
\end{verbatim}.

In this case, \gaiahub{} will try to use the SIMBAD search engine \citep{Simbad} to obtain the central coordinates $(\alpha, \delta)$ of the Sculptor dSph galaxy and its projected size in the sky. These quantities, if available, will define a cone region in the sky where \gaiahub{} will download the \Gaia{} data and try to find suitable \textit{HST} observations. Specifically, the search region will be centered in the object's central coordinates and will have a radius $\mbox{\tt search\_radius} = \max(2 r_{\rm ma}, 2) \degree$, where $r_{\rm ma}$ is the length in degrees of the object's optical major axis. Both sky coordinates and the search radius can be manually set by explicitly including the desired value during the call to \gaiahub. For example,

\begin{verbatim}
$ gaiahub --name "Sculptor dSph" \
--search_radius 1.2
\end{verbatim},

will search for all suitable data in a cone of 1.2 degrees around the central coordinates of the Sculptor dSph galaxy. When providing the coordinates explicitly, these will be used instead of those found in SIMBAD. 
These are defined by the $\mbox{\tt --ra}$ and $\mbox{\tt --dec}$ options: 

\begin{verbatim}
$ gaiahub --name "Sculptor dSph" \
--ra 15.05 --dec -33.84 --search_radius 0.25
\end{verbatim}.

Here, since the coordinates and $\mbox{\tt --search\_radius}$ are explicitly set by the user, the option $\mbox{\tt --name}$ will only be used to create a new folder named as the object and subsequent sub folders where the results and intermediate files will be stored. If no name is provided, the folder will be named "Output".

\subsubsection{Advanced execution}\label{Apx:Advanced_execution}

\gaiahub{} includes a wide range of options that allow the user to fine-tune their search and the way the PMs are computed. Some options are implemented as flags that can be included in the execution call in order to activate a certain feature or behavior. Other options must be followed by a string, number, or list of numbers or strings separated by a space, in order to specify the value to be used. For example,

\begin{verbatim}
$ gaiahub --name "Omega Cent" \
--ra 201.405 --dec -47.667 \
--search_radius 0.1 \
--hst_filters "F814W" "F606W" --use_members \
--use_sat --preselect_cmd \ 
--preselect_pm --use_only_good_gaia
\end{verbatim},

will search for \Gaia{} and \textit{HST} data in a cone of 0.1 degrees around the coordinates (RA, Dec)$ = (201.405, -47.667)$, but only in the F814W and F606W filters for the \textit{HST}. The $\mbox{\tt --use\_members}$ flag forces \gaiahub{} to use only member stars during the alignment between epochs, while $\mbox{\tt --use\_sat}$ allows the use of saturated stars for the same purpose\footnote{\HSTONEPASS{} detects when a star is saturated and tries to reconstruct its position. This provides good results for mildly saturated stars that are not close to the edge of the image ($\Delta r_{HST} \lesssim 1 \mas$). However, this functionality is only available in latest version of \HSTONEPASS, which at the moment of writing this paper, has not yet been made publicly available. The use of the $\mbox{\tt --use\_sat}$ flag will have no effect with the currently available version of \HSTONEPASS{} falling into the default behavior and ignoring saturated sources.}. The $\mbox{\tt --preselect\_cmd}$ and $\mbox{\tt --preselect\_pm}$ allows the user, respectively, to do an interactive manual selection of stars in the CMD and the VPD before the automatic membership selection in the PM space. Lastly, the  $\mbox{\tt --use\_only\_good\_gaia}$ flag forces \gaiahub{} to use only stars that have passed the quality cuts proposed in \citet{Riello2020} and \cite{Lindegren2020} to do the alignment \citep[see Section 2.1.1 from][for more information]{Martinez-Garcia2021}.

The user can also chose to to run \gaiahub{} in a completely automatic, non-interactive way. This is done using the $\mbox{\tt --quiet}$ flag, which forces \gaiahub{} to adopt all default values in case some quantity was not defined during the call to the program. This is useful to execute \gaiahub{} within another script.

\subsubsection{Modular Execution}\label{Apx:Modular_Execution}

\gaiahub{} consist of a main program and a module file, $\mbox{\tt gaiahubmod}$, containing all the required functions and routines for its execution. This module can be added to the Python path and be imported into a Python session. For example the Python script 

\begin{verbatim}
$ import gaiahubmod as gh \
obs, data = gh.search_mast(201.405, -47.667)
\end{verbatim},

will return the tables $\mbox{\tt obs}$ and $\mbox{\tt data}$, containing all suitable \textit{HST} observations around (RA, Dec) $= (201.405^\circ, -47.667^\circ)$.

\subsubsection{Execution times, data download and storage}

The execution time of \gaiahub{} greatly depends on the amount of data being used, the options used, and on whether it is the first execution. Given a particular field, the first execution will normally be the slowest, as \gaiahub{} has to download all the data, run the detection of sources in the \textit{HST} images, and then perform the actual fitting between epochs and compute the PMs. The first two steps are the most time consuming, and in cases where a large amount of \textit{HST} images are being used, \gaiahub{} may need several minutes to download and reduce them. However, these first steps normally have to be performed just once, as \gaiahub{} stores data locally in the computer where it is being executed. Subsequent runs of the script will be much faster, requiring normally less than a minute in a field with 4 \textit{HST} images and few hundreds of stars (as tested on a 2.8 GHz Intel Core i7 with four cores). Fields with several thousands of stars can take up to several minutes depending on the options.

If a new search is made with different coordinates or search radius, \gaiahub{} will then download new \Gaia{} data and \textit{HST} images if necessary. Each \textit{HST} image require from around 200 Megabytes of free space in the disk, which could rapidly increase the total required free space depending on the number of images. 

\section{Membership selection}\label{Apx:Membership_selection}

\gaiahub{} performs automatic membership selection following a slightly modified version of the method described in \citet{delPino2021}. A 2D Gaussian model is fitted to the relative PMs measured by \gaiahub. The PMs and their uncertainties are then Mahalanobis whitened\footnote{A whitening transformation is a linear transformation that transforms a vector of random variables with a known covariance matrix into a set of new variables whose covariance is the identity matrix, meaning that they are uncorrelated and each have variance 1. It can be decomposed in a decorrelation and a standarization of the data.} (ZCA) as:

\begin{equation}
    \mu_{ZCA} = U\Lambda^{-1/2}U^T\mu
\end{equation},

where $\mu$ is the vector of the PMs and their uncertainties, $\Lambda$ are the eigenvalues of the covariance matrix, and $U$ the eigenvectors. Stars not fulfilling

\begin{align}
    \left( \mu_{\alpha\star, ZCA}^2 + \mu_{\delta, ZCA}^2 \right)^{1/2} \leq n {\rm, and}  \\
    \left[ \sigma(\mu_{\alpha\star})_{ZCA}^2 + \sigma(\mu_{\delta})_{ZCA}^2 \right]^{1/2} \leq n
\end{align}

are rejected, where $n$ is the number of $\sigma$ (3 by default). A new Gaussian fit is performed on the remaining stars and the process is repeated until convergence. In cases where the method does not converges, the user can manually select stars in the CMD or in the VPD prior to their automatic selection in the PMs space.

\section{Gaia positional uncertainties}

There is not much information about how accurate \Gaia{} positional errors are. However, many studies have reported uncertainties in parallaxes and PMs to be underestimated. By default, \gaiahub{} tries to correct this by multiplying \Gaia{} positional uncertainties by a factor 1.05 and 1.22 for 5-parameter and 6-parameter
solutions, respectively \citep[see Figure 21 in][]{Fabricius2021}. This behavior can be avoided by using the $\mbox{\tt --no\_error\_correction}$, which will force \gaiahub{} to use the nominal positional errors listed in the $\mbox{gaia\_source}$ table.

\bibliographystyle{aasjournal}
\bibliography{GaiaHub.bib}

\end{document}